\documentclass[12pt]{article}
\usepackage{amsfonts}

\usepackage{psfig}
\usepackage{amsmath}
\usepackage{latexsym}
\newtheorem{teo}{Theorem}[section]
\newtheorem{defin}{Definition}[section]
\begin{document}

\title{A probabilistic operator symbol framework for quantum information}
\author{M. A. Man'ko\thanks{%
P. N. Lebedev Physical Institute, Leninskii Prospect 53, Moscow 119991,
Russia, e-mail: mmanko@sci.lebedev.ru}, V. I. Man'ko\thanks{%
P. N. Lebedev Physical Institute, Leninskii Prospect 53, Moscow 119991,
Russia, e-mail: manko@sci.lebedev.ru}\quad and R. Vilela Mendes\thanks{%
CMAF, Complexo Interdisciplinar, Universidade de Lisboa, Av. Gama Pinto, 2 -
P1699 Lisboa Codex, Portugal, e-mail: vilela@cii.fc.ul.pt;
http://label2.ist.utl.pt/vilela/}}
\date{}
\maketitle

\begin{abstract}
Hilbert space operators may be mapped onto a space of ordinary functions
(operator symbols) equipped with an associative (but noncommutative)
star-product. A unified framework for such maps is reviewed. Because of its
clear probabilistic interpretation, a particular class of operator symbols
(tomograms) is proposed as a framework for quantum information problems.
Qudit states are identified with maps of the unitary group into the simplex.
The image of the unitary group on the simplex provides a geometrical
characterization of the nature of the quantum states. Generalized
measurements, typical quantum channels, entropies and entropy inequalities
are discussed in this setting.
\end{abstract}

\section{Introduction}

Algebras of Hilbert space operators may be mapped onto algebras of ordinary
functions on linear spaces, with an associative but non-commutative \textit{%
star product} (see, e.g. \cite{Straton,Fronsdal}). The images of the Hilbert
space operators are called \textit{operator symbols.} Weyl maps \cite
{Wigner32,Moyal}, $s$-ordered operator symbols \cite{Cahill}, their partial
cases \cite{Glauber63,Sudarshan,Husimi,Kano} and tomograms \cite
{Manko3,DodPLA,OlgaJETP} are examples of this correspondence between Hilbert
space operator algebras and function algebras \cite{Olga1,MarmoPhysScr}. In
the case of tomograms, the operator symbols of the density operators of
quantum mechanics are families of ordinary probability distributions \cite
{Mancini1,Mancini2,Olga1,Olga2}.

A unified framework for operator symbols is presented in Sect. 2 and their
main properties are reviewed. Many of the results in Sects. 2 and 3 are
scattered in previous publications and are collected here to make the paper
reasonably self-contained.

Finite dimensional systems (spin tomograms) are studied in Sect. 3. These
operator symbols are then proposed as a framework for quantum information
problems. Qudit states are identified with maps of the unitary group into
the simplex. The image of the unitary group on the simplex provides a
geometrical characterization of the nature of the quantum states. In the
remaining sections, generalized measurements, typical quantum channels,
entropies and entropy inequalities are discussed in this setting.

\section{Operator symbols for quantum mechanical observables}

In quantum mechanics, observables are selfadjoint operators acting on the
Hilbert space of states $H$. We map operators onto functions in a vector
space $X$ in the following way:

Given the Hilbert space $H$ and a trace-class operator $\hat{A}$ acting on
this space, let $\hat{U}(\mathbf{x})$ be a family of operators on $H$,
labelled by vectors $\mathbf{x\in X}$. We construct the $c$-number function $%
\left\{ f_{\hat{A}}(\mathbf{x}):X\rightarrow \mathbb{C}\right\} $ (and call it
the \textit{symbol of the operator} $\hat{A}$) by 
\begin{equation}
f_{\hat{A}}(\mathbf{x})=\mbox{Tr}\left[ \hat{A}\hat{U}(\mathbf{x})\right] .
\label{eq.A1}
\end{equation}
Let us suppose that relation~(\ref{eq.A1}) has an inverse, i.e., there is a
set of operators $\hat{D}(\mathbf{x})$ acting on the Hilbert space such that 
\begin{equation}
\hat{A}=\int_{X}f_{\hat{A}}(\mathbf{x})\hat{D}(\mathbf{x})~d\mathbf{x}%
,\qquad \mbox{Tr}\,\hat{A}=\int_{X}f_{\hat{A}}(\mathbf{x})\,\mbox{Tr}\,\hat{D%
}(\mathbf{x})~d\mathbf{x}.  \label{eq.A2}
\end{equation}
Equations (\ref{eq.A1}) and~(\ref{eq.A2}) define an invertible map from the
operator $\hat{A}$ onto the function $f_{\hat{A}}(\mathbf{x})$. Multiplying
both sides of Eq.(\ref{eq.A2}) by the operator $\hat{U}(\mathbf{x}^{\prime
}) $ and taking the trace, one obtains a consistency condition for the
operators $\hat{U}(\mathbf{x}^{\prime })$ and $\hat{D}(\mathbf{x})$ 
\[
\mbox{Tr}\left[ \hat{D}(\mathbf{x})\hat{U}(\mathbf{x}^{\prime })\right]
=\delta \left( \mathbf{x-x}^{\prime }\right) . 
\]

For two functions $f_{\hat{A}}(\mathbf{x})$ and $f_{\hat{B}}(\mathbf{x})$,
corresponding to two operators $\hat{A}$ and $\hat{B}$, a star-product is
defined by 
\begin{equation}
f_{\hat{A}\hat{B}}(\mathbf{x})=f_{\hat{A}}(\mathbf{x})*f_{\hat{B}}(\mathbf{x}%
):=\mbox{Tr}\left[ \hat{A}\hat{B}\hat{U}(\mathbf{x})\right] .  \label{eq.A5}
\end{equation}
Since the standard product of operators on a Hilbert space is associative,
Eq. (\ref{eq.A5}) also defines an associative product for the functions $f_{%
\hat{A}}(\mathbf{x})$, i.e., 
\begin{equation}
f_{\hat{A}}(\mathbf{x})*\Big(f_{\hat{B}}(\mathbf{x})*f_{\hat{C}}(\mathbf{x})%
\Big)=\Big(f_{\hat{A}}(\mathbf{x})*f_{\hat{B}}(\mathbf{x})\Big) *f_{\hat{C}}(%
\mathbf{x}).  \label{eq.A6}
\end{equation}

Let us suppose that there is another map, analogous to the one in (\ref
{eq.A1}) and~(\ref{eq.A2}), defined by the operator families $\hat{U}_{1}(%
\mathbf{y})$ and $\hat{D}_{1}(\mathbf{y})$. Then one has 
\begin{equation}
\phi _{\hat{A}}(\mathbf{y})=\mbox{Tr}\left[ \hat{A}\hat{U}_{1}(\mathbf{y}%
)\right]  \label{eq.A21}
\end{equation}
and the inverse relation 
\begin{equation}
\hat{A}=\int \phi _{A}(\mathbf{y})\hat{D}_{1}(\mathbf{y})~d\mathbf{y}.
\label{eq.A22}
\end{equation}
The function $f_{\hat{A}}(\mathbf{x})$ will be related to the function $\phi
_{\hat{A}}(\mathbf{y})$ by 
\begin{equation}
\phi _{\hat{A}}(\mathbf{y})=\int f_{\hat{A}}(\mathbf{x})\,\mbox{Tr}\left[ 
\hat{D}(\mathbf{x})\hat{U}_{1}(\mathbf{y})\right] \,d\mathbf{x}
\label{eq.A23}
\end{equation}
with the inverse relation 
\begin{equation}
f_{\hat{A}}(\mathbf{x})=\int \phi _{\hat{A}}(\mathbf{y})\,\mbox{Tr}\left[ 
\hat{D}_{1}(\mathbf{y})\hat{U}(\mathbf{x})\right] \,d\mathbf{y}.
\label{eq.A24}
\end{equation}
The functions $f_{\hat{A}}(\mathbf{x})$ and $\phi _{\hat{A}}(\mathbf{y})$,
corresponding to different maps, are connected by the invertible integral
transform given by Eqs.(\ref{eq.A23}) and (\ref{eq.A24}) with the
intertwining kernels 
\begin{equation}
K_{1}(\mathbf{x},\mathbf{y})=\mbox{Tr}\,\Big[\hat{D}(\mathbf{x})\hat{U}_{1}(%
\mathbf{y})\Big ]  \label{eq.A28aa}
\end{equation}
and 
\begin{equation}
K_{2}(\mathbf{x},\mathbf{y})=\mbox{Tr}\,\Big[\hat{D}_{1}(\mathbf{y})\hat{U}(%
\mathbf{x})\Big].  \label{eq.A28bb}
\end{equation}

Using formulae~(\ref{eq.A1}) and (\ref{eq.A2}), one writes a composition
rule for two symbols $f_{\hat{A}}(\mathbf{x})$ and $f_{\hat{B}}(\mathbf{x})$%
determining their star-product 
\begin{equation}
f_{\hat{A}}(\mathbf{x})*f_{\hat{B}}(\mathbf{x})=\int f_{\hat{A}}(\mathbf{x}%
^{\prime \prime })f_{\hat{B}}(\mathbf{x}^{\prime })K(\mathbf{x}^{\prime
\prime },\mathbf{x}^{\prime },\mathbf{x})\,d\mathbf{x}^{\prime }\,d\mathbf{x}%
^{\prime \prime }.  \label{eq.A25}
\end{equation}
The kernel in (\ref{eq.A25}) is determined by the trace of the product of
the operators used to construct the map 
\begin{equation}
K(\mathbf{x}^{\prime \prime },\mathbf{x}^{\prime },\mathbf{x})=\mbox{Tr}%
\left[ \hat{D}(\mathbf{x}^{\prime \prime })\hat{D}(\mathbf{x}^{\prime })\hat{%
U}(\mathbf{x})\right] .  \label{eq.A26}
\end{equation}
Equation (\ref{eq.A26}) can be extended to the case of the star-product of $%
N $ symbols of operators $\hat{A}_{1},\hat{A}_{2},\ldots ,\hat{A}_{N}$ 
\begin{eqnarray}
f_{\hat{A}_{1}}(\mathbf{x})*f_{\hat{A}_{2}}(\mathbf{x})*\cdots *f_{\hat{A}%
_{N}}(\mathbf{x}) &=&\int f_{\hat{A}_{1}}(\mathbf{x}_{1})f_{\hat{A}_{2}}(%
\mathbf{x}_{2})\cdots f_{\hat{A}_{N}}(\mathbf{x}_{N})  \label{eq.A26'} \\
&&\times K\left( \mathbf{x}_{1},\mathbf{x}_{2},\ldots ,\mathbf{x}_{N},%
\mathbf{x}\right) \,d\mathbf{x}_{1}\,d\mathbf{x}_{2}\cdots \,d\mathbf{x}_{N}
\nonumber
\end{eqnarray}
with kernel 
\begin{equation}
K\left( \mathbf{x}_{1},\mathbf{x}_{2},\ldots ,\mathbf{x}_{N},\mathbf{x}%
\right) =\mbox{Tr}\left[ \hat{D}(\mathbf{x}_{1})\hat{D}(\mathbf{x}%
_{2})\cdots \hat{D}(\mathbf{x}_{N})\hat{U}(\mathbf{x})\right] .
\label{eq.A26''}
\end{equation}
The trace of an operator $\hat{A}^{N}$ is determined by 
\begin{eqnarray}
&&\mbox{Tr}\,\hat{A}^{N}=\int f_{\hat{A}}(\mathbf{x}_{1})f_{\hat{A}}(\mathbf{%
x}_{2})\cdots f_{\hat{A}}(\mathbf{x}_{N})  \nonumber  \label{eq.A26'''} \\
&&\times \mbox{Tr}\left[ \hat{D}(\mathbf{x}_{1})\hat{D}(\mathbf{x}%
_{2})\cdots \hat{D}(\mathbf{x}_{N})\right] \,d\mathbf{x}_{1}\,d\mathbf{x}%
_{2}\cdots \,d\mathbf{x}_{N}.
\end{eqnarray}

Consider now a linear superoperator $L$ acting in linear space of operators.
The map of operators $\hat{A}\rightarrow L\hat{A}$ induces a corresponding
map of their symbols 
\begin{equation}
f_{\hat{A}}(\mathbf{x})\rightarrow \hat{L}f_{\hat{A}}(\mathbf{x})=f_{L\hat{A}%
}(\mathbf{x}).  \label{AR1}
\end{equation}
The integral form of this map 
\begin{equation}
\hat{L}f_{\hat{A}}(\mathbf{x})=\int \Pi _{L}(\mathbf{x},\mathbf{x}^{\prime
})f_{\hat{A}}(\mathbf{x}^{\prime })\,d\mathbf{x}^{\prime }  \label{AR2}
\end{equation}
is determined by the kernel 
\begin{equation}
\Pi _{L}(\mathbf{x},\mathbf{x}^{\prime })=\mbox{Tr}\left[ \hat{U}(\mathbf{x})%
\Big(L\hat{D}(\mathbf{x}^{\prime })\Big)\right] .  \label{AR3}
\end{equation}

\subsection{The Weyl operator symbols}

As operator family $\hat{U}(\mathbf{x})$, we take the Fourier transform of
the displacement operator $d(\xi )$ 
\begin{equation}
\hat{U}(\mathbf{x})=\int \exp \left( \frac{x_{1}+ix_{2}}{\sqrt{2}}%
\mbox{\boldmath$\xi$}^{*}-\frac{x_{1}-ix_{2}}{\sqrt{2}}\mbox{\boldmath$\xi$}%
\right) d(\mbox{\boldmath$\xi$})\pi ^{-1}~d^{2}\mbox{\boldmath$\xi$},
\label{eq.A27}
\end{equation}
where $\mbox{\boldmath$\xi$}$ is a complex number, $\mbox{\boldmath$\xi$}%
=\xi _{1}+i\xi _{2}$, and the vector $\mathbf{x}=(x_{1},x_{2})$ may be
interpreted as $\mathbf{x}=(q,p)$, with $q$ and $p$ being the position and
momentum. One sees that $\mbox{Tr}\,\hat{U}(\mathbf{x})=1$. The displacement
operator (creating coherent states from the vacuum) may be expressed through
creation and annihilation operators in the form 
\begin{equation}
d(\mbox{\boldmath$\xi$})=\exp (\mbox{\boldmath$\xi$}\hat{a}^{\dagger }-%
\mbox{\boldmath$\xi$}^{*}\hat{a}),  \label{eq.A28}
\end{equation}
\begin{equation}
\hat{a}=\frac{\hat{q}+i\hat{p}}{\sqrt{2}},\qquad \hat{a}^{\dagger }=\frac{%
\hat{q}-i\hat{p}}{\sqrt{2}}.  \label{eq.A29}
\end{equation}
The operator $\hat{a}$ and its Hermitian conjugate $\hat{a}^{\dagger }$
satisfy the boson commutation relation $[\hat{a},\hat{a}^{\dagger }]=\hat{%
\mathbf{1}}.$

The Weyl symbol for an operator $\hat{A}$ reads 
\begin{equation}
W_{\hat{A}}(\mathbf{x})=\mbox{Tr}\left[ \hat{A}\hat{U}(\mathbf{x})\right] ,
\label{eq.A30}
\end{equation}
$\hat{U}(\mathbf{x})$ being given by Eq.(\ref{eq.A27}). The Weyl symbols of
the identity operator $\hat{\mathbf{1}}$, the position operator $\hat{q}$
and the momentum operator $\hat{p}$ are 
\begin{equation}
W_{\hat{\mathbf{1}}}(q,p)=1\qquad W_{\hat{q}}(q,p)=q\qquad W_{\hat{p}%
}(q,p)=p.  \label{eq.A30'}
\end{equation}
The inverse transform, which expresses the operator $\hat{A}$ through its
Weyl symbol, is 
\begin{equation}
\hat{A}=\int W_{\hat{A}}(\mathbf{x})\hat{U}(\mathbf{x})\,\frac{d\mathbf{x}}{%
2\pi }\,.  \label{eq.A31}
\end{equation}
That is, the operator $\hat{D}(\mathbf{x})$ in formula~(\ref{eq.A2}) is
related to $\hat{U}(\mathbf{x})$ by 
\begin{equation}
\hat{D}(\mathbf{x})=\frac{\hat{U}(\mathbf{x})}{2\pi }\,.  \label{eq.A32}
\end{equation}
The star-product of the Weyl symbols of two operators $\hat{A}_{1}$ and $%
\hat{A}_{2}$ , expressed through Weyl symbols by 
\begin{equation}
\hat{A}_{1}=\int W_{\hat{A}_{1}}(\mathbf{x}^{\prime })\hat{U}(\mathbf{x}%
^{\prime })\,\frac{d\mathbf{x}^{\prime }}{2\pi },\qquad \hat{A}_{2}=\int W_{%
\hat{A}_{2}}(\mathbf{x}^{\prime \prime })\hat{U}(\mathbf{x}^{\prime \prime
})\,\frac{d\mathbf{x}^{\prime \prime }}{2\pi },  \label{eq.A33}
\end{equation}
with vectors $\mathbf{x}^{\prime }=(x_{1}^{\prime },x_{2}^{\prime })$ and $%
\mathbf{x}^{\prime \prime }=(x_{1}^{\prime \prime },x_{2}^{\prime \prime })$%
, is the operator $\hat{A}$ with Weyl symbol 
\begin{eqnarray}
&&W_{\hat{A}}(\mathbf{x})=W_{\hat{A}_{1}}(\mathbf{x})*W_{\hat{A}_{2}}(%
\mathbf{x})=\int \frac{d\mathbf{x}^{\prime }\,d\mathbf{x}^{\prime \prime }}{%
\pi ^{2}}\,W_{\hat{A}_{1}}(\mathbf{x}^{\prime })W_{\hat{A}_{2}}(\mathbf{x}%
^{\prime \prime })  \nonumber \\
&&\times \exp \Big\{2i\Big[(x_{2}^{\prime }-x_{2})(x_{1}-x_{1}^{\prime
\prime })+(x_{1}^{\prime }-x_{1})(x_{2}^{\prime \prime }-x_{2})\Big]\Big\}.
\label{eq.A36}
\end{eqnarray}

\subsection{The $s$-ordered operator symbols}

The $s$-ordered symbol \cite{Cahill} $W_{\hat{A}}(\mathbf{x},s)$ of the
operator $\hat{A}$ is 
\begin{equation}
W_{\hat{A}}(\mathbf{x},s)=\mbox{Tr}\left[ \hat{A}\hat{U}(\mathbf{x}%
,s)\right],  \label{s1}
\end{equation}
with a real parameter $s$, real vector $\mathbf{x}=(x_{1},x_{2})$ and
operator $\hat{U}(\mathbf{x},s)$ 
\begin{equation}
\hat{U}(\mathbf{x},s)=\frac{2}{1-s}\,d(\alpha _{\mathbf{x}})\,q^{\hat{a}%
^{\dagger }\hat{a}}(s)\,d(-\alpha _{\mathbf{x}}),  \label{s2}
\end{equation}
the displacement operator being 
\begin{equation}
d(\alpha _{\mathbf{x}})=\exp \left( \alpha _{\mathbf{x}}\hat{a}^{\dagger
}-\alpha _{\mathbf{x}}^{*}\hat{a}\right) .  \label{s3}
\end{equation}
Also 
\begin{equation}
\alpha _{\mathbf{x}}=x_{1}+ix_{2},\qquad \alpha _{\mathbf{x}%
}^{*}=x_{1}-ix_{2},\qquad x_{1}=\frac{q}{\sqrt{2}},\qquad x_{2}=\frac{p}{%
\sqrt{2}}  \label{s4}
\end{equation}
and 
\begin{equation}
q(s)=\frac{s+1}{s-1}\,.  \label{s5}
\end{equation}
The coefficient in Eq.~(\ref{s2}) leads to $\mbox{Tr}\left[ \hat{U}(\mathbf{x%
},s)\right] =1$, meaning that the symbol of the identity operator equals $1$.

The operator $\hat{A}$ is obtained from 
\begin{equation}
\hat{A}=\frac{1}{\pi }\,\frac{1+s}{1-s}\,\int W_{\hat{A}}(\mathbf{x},s)\hat{U%
}(\mathbf{x},-s)\,d(\mathbf{x}).  \label{s10}
\end{equation}
This means that, for $s$-ordered symbols, the operator $\hat{D}(\mathbf{x})$
in the general formula~(\ref{eq.A2}) takes the form 
\begin{equation}
\hat{D}(\mathbf{x})=\frac{1}{\pi }\,\frac{1+s}{1-s}\,\hat{U}(\mathbf{x},-s).
\label{s11}
\end{equation}
If $\hat{A}$ is a density operator $\hat{\rho}$ \cite
{landau27,vonneumann27,vonneumann-book32}, for the values of the parameters $%
s=0,1,-1$, the corresponding symbols are respectively the Wigner,
Glauber--Sudarshan and Husimi quasidistributions.

For the explicit form of the kernel for the product of $N$ operator symbols
we refer to \cite{Olga1}.

\subsection{The tomographic operator symbols}

Density operators may be mapped onto probability distribution functions
(tomograms) of one random variable $X$ and two real parameters $\mu $ and $%
\nu $. This map has been used to provide a formulation of quantum mechanics,
in which quantum states are described by a parametrized family of
probability distributions \cite{Mancini1,Mancini2}, alternative to the
description of the states by wave functions or density operators. The
tomographic map has been used to reconstruct the quantum state, to obtain
the Wigner function by measuring the state tomogram, to define quantum
characteristic exponents \cite{MendesPhysica} and for the simulation of
nonstationary quantum systems \cite{Arkhipov}.

Here we discuss the tomographic map as an example of the general operator
symbol framework. The operator $\hat{A}$ is mapped onto the function $f_{%
\hat{A}}(\mathbf{x})$, where $\mathbf{x}\equiv (X,\mu ,\nu )$, which we
denote as $w_{\hat{A}}(X,\mu ,\nu )$ depending on the coordinate $X$ and the
reference frame parameters $\mu $ and $\nu $ 
\begin{equation}
w_{\hat{A}}(X,\mu ,\nu )=\mbox{Tr}\left[ \hat{A}\hat{U}(\mathbf{x})\right] .
\label{eq.53}
\end{equation}
The function $w_{\hat{A}}(X,\mu ,\nu )$ is the symbol of the operator $\hat{A%
}$. The operator $\hat{U}(x)$ is 
\begin{eqnarray}
\hat{U}(\mathbf{x}) &\equiv &\hat{U}(X,\mu ,\nu )=\exp \left( \frac{i\lambda 
}{2}\left( \hat{q}\hat{p}+\hat{p}\hat{q}\right) \right) \exp \left( \frac{%
i\theta }{2}\left( \hat{q}^{2}+\hat{p}^{2}\right) \right) \mid X\rangle
\langle X\mid  \nonumber \\
&&\qquad ~\times \exp \left( -\frac{i\theta }{2}\left( \hat{q}^{2}+\hat{p}%
^{2}\right) \right) \exp \left( -\frac{i\lambda }{2}\left( \hat{q}\hat{p}+%
\hat{p}\hat{q}\right) \right)  \nonumber \\
\qquad &=&\hat{U}_{\mu \nu }\mid X\rangle \langle X\mid \hat{U}_{\mu \nu
}^{\dagger },  \label{eq.54}
\end{eqnarray}
where $\hat{q}$ and $\hat{p}$ are position and momentum operators and the
angle $\theta $ and parameter $\lambda $ are related to the reference frame
parameters by 
\[
\mu =e^{\lambda }\cos \theta \qquad \nu =e^{-\lambda }\sin \theta . 
\]
Moreover, 
\begin{equation}
\hat{X}\mid X\rangle =X\mid X\rangle  \label{eq.54'}
\end{equation}
and $\mid X\rangle \langle X\mid $ is a projection density. One has the
canonical transform of quadratures 
\[
\hat{X}=\hat{U}_{\mu \nu }\,\hat{q}\,\hat{U}_{\mu \nu }^{\dagger }=\mu \hat{q%
}+\nu \hat{p}, 
\]
\[
\hat{P}=\hat{U}_{\mu \nu }\,\hat{p}\,\hat{U}_{\mu \nu }^{\dagger }=\frac{1+%
\sqrt{1-4\mu ^{2}\nu ^{2}}}{2\mu }\,\hat{p}-\frac{1-\sqrt{1-4\mu ^{2}\nu ^{2}%
}}{2\nu }\,\hat{q}. 
\]

Using the approach of \cite{Manko1} one obtains the relation 
\[
\hat{U}(X,\mu ,\nu )=\delta (X-\mu \hat{q}-\nu \hat{p}). 
\]
In the case we are considering, the inverse transform determining the
operator in terms of the tomogram symbol will be of the form 
\begin{equation}
\hat{A}=\int w_{\hat{A}}(X,\mu ,\nu )\hat{D}(X,\mu ,\nu )\,dX\,d\mu \,d\nu,
\label{eq.55}
\end{equation}
where~\cite{Manko2,Manko3} 
\begin{equation}
\hat{D}(\mathbf{x})\equiv \hat{D}(X,\mu ,\nu )=\frac{1}{2\pi }\exp \left(
iX-i\nu \hat{p}-i\mu \hat{q}\right),  \label{eq.56}
\end{equation}
i.e., 
\begin{equation}
\hat{D}(X,\mu ,\nu )=\frac{1}{2\pi }\exp (iX)d\Big(\mbox{\boldmath$\xi$}(\mu
,\nu )\Big).  \label{eq.56q}
\end{equation}
The unitary displacement operator in (\ref{eq.56q}) now reads 
\[
d\Big(
\mbox{\boldmath$\xi$}(\mu ,\nu )\Big)
=\exp \Big(\mbox{\boldmath$\xi$}(\mu ,\nu )\hat{a}^{+}-{\mbox{\boldmath$\xi$}%
}^{*}(\mu ,\nu )\hat{a}\Big), 
\]
where $\mbox{\boldmath$\xi$}(\mu ,\nu )=\xi _{1}+i\xi _{2}$ with $\xi _{1}=%
\mbox{Re}\,(\mbox{\boldmath$\xi$})={\nu }/{\sqrt{2}}$ and $\xi _{2}=\mbox{Im}%
\,(\mbox{\boldmath$\xi$})=-{\mu }/{\sqrt{2}}$.

The trace of the above operator provides the kernel determining the trace of
an arbitrary operator in the tomographic representation 
\[
\mbox{Tr}\,\hat{D}(\mathbf{x})=e^{iX}\delta (\mu )\delta (\nu ). 
\]
The operators $a^{\dagger }$ and $a$ are creation and annihilation
operators. The function $w_{\hat{A}}(X,\mu ,\nu )$ satisfies the relation 
\begin{equation}
w_{\hat{A}}\left( \lambda X,\lambda \mu ,\lambda \nu \right) =\frac{1}{%
|\lambda |}\,w_{\hat{A}}(X,\mu ,\nu )  \label{eq.56'}
\end{equation}
meaning that the tomographic symbols are homogeneous functions of three
variables.

For the density operator of a pure state $\mid \psi \rangle \langle \psi
\mid $, the tomographic symbol reads \cite{MendesPLA} 
\begin{equation}
w_{\psi }(X,\mu ,\nu )=\frac{1}{2\pi |\nu |}\left| \int \psi (y)\exp \left( 
\frac{i\mu }{2\nu }y^{2}-\frac{iX}{\nu }y\right) dy\right| ^{2}.  \label{TS}
\end{equation}

If one takes two operators $\hat{A}_{1}$ and $\hat{A}_{2}$ 
\begin{eqnarray}
\hat{A}_{1} &=&\int w_{\hat{A}_{1}}(X^{\prime },\mu ^{\prime },\nu ^{\prime
})\hat{D}(X^{\prime },\mu ^{\prime },\nu ^{\prime })\,dX^{\prime }\,d\mu
^{\prime }\,d\nu ^{\prime },  \nonumber \\
&&  \label{eq.57} \\
\hat{A}_{2} &=&\int w_{\hat{A}_{2}}(X^{\prime \prime },\mu ^{\prime \prime
},\nu ^{\prime \prime })\hat{D}(X^{\prime \prime },\mu ^{\prime \prime },\nu
^{\prime \prime })dX^{\prime \prime }\,d\mu ^{\prime \prime }\,d\nu ^{\prime
\prime },  \nonumber
\end{eqnarray}
the tomographic symbol of the product $\hat{A}=\hat{A}_{1}\hat{A}_{2}$ is
the star-product 
\[
w_{\hat{A}}(X,\mu ,\nu )=w_{\hat{A}_{1}}(X,\mu ,\nu )*w_{\hat{A}_{2}}(X,\mu
,\nu ), 
\]
that is, 
\begin{equation}
w_{\hat{A}}(X,\mu ,\nu )=\int w_{\hat{A}_{1}}(\mathbf{x}^{\prime \prime })w_{%
\hat{A}_{2}}(\mathbf{x}^{\prime })K(\mathbf{x}^{\prime \prime },\mathbf{x}%
^{\prime },\mathbf{x})\,d\mathbf{x^{\prime \prime }}\,d\mathbf{x^{\prime }},
\label{eq.58}
\end{equation}
with kernel given by 
\begin{equation}
K(\mathbf{x}^{\prime \prime },\mathbf{x}^{\prime },\mathbf{x})=\mbox{Tr}%
\left[ \hat{D}(X^{\prime \prime },\mu ^{\prime \prime },\nu ^{\prime \prime
})\hat{D}(X^{\prime },\mu ^{\prime },\nu ^{\prime })\hat{U}(X,\mu ,\nu
)\right] .  \label{eq.59}
\end{equation}
The explicit form of the kernel reads 
\begin{eqnarray}
&&K(X_{1},\mu _{1},\nu _{1},X_{2},\mu _{2},\nu _{2},X,\mu ,\nu )  \nonumber
\\
&=&\frac{\delta \Big(\mu (\nu _{1}+\nu _{2})-\nu (\mu _{1}+\mu _{2})\Big)}{%
4\pi ^{2}}\,\exp \left( \frac{i}{2}\Big\{\left( \nu _{1}\mu _{2}-\nu _{2}\mu
_{1}\right) +2X_{1}+2X_{2}\right.  \nonumber \\
&&\left. \left. -\left[ \frac{1}{\nu }\left( \nu _{1}+\nu _{2}\right) +\frac{%
1}{\mu }\left( \mu _{1}+\mu _{2}\right) \right] X\right\} \right) ,
\label{KERNEL}
\end{eqnarray}
and the kernel for the star-product of $N$ operators is 
\begin{eqnarray}
&&K\left( X_{1},\mu _{1},\nu _{1},X_{2},\mu _{2},\nu _{2},\ldots ,X_{N},\mu
_{N},\nu _{N},X,\mu ,\nu \right)  \nonumber \\
&=&\frac{\delta \left( \mu \sum_{j=1}^{N}\nu _{j}-\nu \sum_{j=1}^{N}\mu
_{j}\right) }{(2\pi )^{N}}\,\exp \left( \frac{i}{2}\,\left\{
\sum_{k<j=1}^{N}\left( \nu _{k}\mu _{j}-\nu _{j}\mu _{k}\right)
+2\sum_{j=1}^{N}X_{j}\right. \right.  \nonumber \\
&&\left. \left. -\left[ \frac{1}{\nu }\left( \sum_{j=1}^{N}\nu _{j}\right) +%
\frac{1}{\mu }\left( \sum_{j=1}^{N}\mu _{j}\right) \right] X\right\} \right)
.  \label{KERNELSTAR}
\end{eqnarray}

\section{Operator symbols for spin systems}

Of particular importance for quantum information purposes are
finite-dimensional spin systems (qubits, qutrits, etc.). Therefore, we
describe here the tomographic operator symbols for spin systems. Further
details may be obtained from Refs. \cite
{DodPLA,OlgaJETP,OlgaJRLRspin,MarmoPhysScr,Klimov,Casta}. In this case, the
physical interpretation of the symbol is as the set of measurable mean
values of the operator in a state with a given spin projection in a rotated
reference frame.

\subsection{Review of spin state properties and spin-related operators}

To set the notation, we describe here some standard operators used to
discuss the properties of spin states. For arbitrary values of spin, let the
observable $\hat{A}^{(j)}$ be represented by a matrix in the standard basis
of angular momentum generators $\hat{J}_{i}$, $i=1,\,2,\,3$, 
\begin{equation}
\hat{J}^{2}\mid jm\rangle =j(j+1)\mid jm\rangle, \qquad \hat{J}_{3}\mid
jm\rangle =m\mid jm\rangle  \label{eq.1}
\end{equation}
as 
\begin{equation}
\hat{A}^{(j)}=\sum_{m=-j}^{j}\sum_{m^{\prime }=-j}^{j}A_{mm^{\prime
}}^{(j)}\mid jm\rangle \langle jm^{\prime }\mid,  \label{eq.2}
\end{equation}
where 
\begin{equation}
A_{mm^{\prime }}^{(j)}=\langle jm\mid \hat{A}^{(j)}\mid jm^{\prime }\rangle.
\qquad m=-j,-j+1,\ldots ,j-1,j .  \label{d1}
\end{equation}

The spin $j$ projector onto the $m_{1}$ component along $z$-axis is denoted 
\begin{equation}
\hat{\Pi}_{m_{1}}^{(j)}=|jm_{1}\rangle \langle jm_{1}|,  \label{a01}
\end{equation}
and the same projector in a reference frame rotated by an element $g$ of $%
SU(2)$ is 
\begin{equation}
\hat{\Pi}_{m_{1}}^{(j)}(g)=R^{\dagger }(g)\,\hat{\Pi}_{m_{1}}^{(j)}\,R(g),
\label{a02}
\end{equation}
$R(g)$ being a rotation operator of the $SU(2)$ irreducible representation
with spin $j$. Since the projectors play an important role in constructing
the tomographic map, we present several different expressions for these
operators. The projector can be given an alternative form in terms of the
Dirac delta-function 
\begin{equation}
\hat{\Pi}_{m_{1}}^{(j)}=\delta \left( m_{1}-\hat{J}_{3}\right) ,  \label{a03}
\end{equation}
and for the rotated projector 
\begin{equation}
\hat{\Pi}_{m_{1}}^{(j)}(g)=\delta \left( m_{1}-R^{\dagger }(g)\hat{J}%
_{3}R(g)\right) ,  \label{a04}
\end{equation}
or, in integral form 
\begin{equation}
\hat{\Pi}_{m_{1}}^{(j)}(g)=\frac{1}{2\pi }\int_{0}^{2\pi }\exp \left[
i\left( m_{1}-R^{\dagger }(g)\hat{J}_{3}R(g)\right) \,\varphi \right]
\,d\varphi \,.  \label{a05}
\end{equation}
Another form of the rotated projector is 
\begin{equation}
\hat{\Pi}_{m_{1}}^{(j)}(g)=\sum_{m_{1}^{\prime }m_{2}^{\prime
}}\,D_{m_{1}m_{2}^{\prime }}^{(j)\,*}(\alpha ,\beta ,\gamma
)\,D_{m_{1}m_{1}^{\prime }}^{(j)}(\alpha ,\beta ,\gamma )\,|jm_{2}^{\prime
}\rangle \langle jm_{1}^{\prime }|\ .  \label{eq09}
\end{equation}
The matrix elements $D_{m_{1}\,m_{1}^{\prime }}^{(j)}\left( \alpha ,\beta
,\gamma \right) $~(Wigner $D$-functions) are the matrix elements of the
operator 
\begin{equation}
R(g)=e^{-i\alpha \hat{J}_{3}}e^{-i\beta \hat{J}_{2}}e^{-i\gamma \hat{J}_{3}}
\end{equation}
of $SU(2)$ group representation ($g$ is an element of the $SU(2)$ group
parametrized by Euler angles). The matrix elements have the explicit form 
\begin{equation}
D_{m^{\prime }m}^{(j)}(\alpha ,\beta ,\gamma )=e^{-im^{\prime }\alpha
}\,d_{m^{\prime }m}^{(j)}(\beta )\,e^{-im\gamma }  \label{eq.6}
\end{equation}
with 
\begin{eqnarray}
d_{m^{\prime }\,m}^{(j)}(\beta ) &=&\sum_{s}\frac{(-1)^{s}\sqrt{%
(j+m)!(j-m)!(j+m^{\prime })!(j-m^{\prime })!}}{s!(j-m^{\prime
}-s)!(j+m-s)!(m^{\prime }-m+s)!}  \nonumber \\
&&\times \left( \cos \frac{\beta }{2}\right) ^{2j+m-m^{\prime }-2s}\left(
-\sin \frac{\beta }{2}\right) ^{m^{\prime }-m+2s}.  \label{drot}
\end{eqnarray}
It is convenient to introduce the irreducible tensor operator for the $SU(2)$
group 
\begin{equation}
\hat{T}_{LM}^{(j)}=\,\sum_{m_{1},m_{2}=-j}^{j}\,(-1)^{j-m_{1}}\,\langle
jm_{2};j-m_{1}|LM\rangle \,|jm_{2}\rangle \,\langle jm_{1}|\ .  \label{TLM}
\end{equation}

The irreducible tensors have the properties (see \cite{Varshalovich}) 
\begin{eqnarray}
&&\hbox{Tr}\left( \hat{T}_{L_{1}M_{1}}^{(j)\,\dagger }\,\hat{T}%
_{L_{2}M_{2}}^{(j)}\right) =\delta _{L_{1}L_{2}}\,\delta _{M_{1}M_{2}},
\label{A} \\
&&\hbox{Tr}\left( \hat{T}_{L_{1}M_{1}}^{(j)}\,\hat{T}_{L_{2}M_{2}}^{(j)}\,%
\hat{T}_{LM}^{(j)}\right) =(-1)^{L_{1}+L_{2}+L-2j}\left( 
\begin{array}{ccc}
L_{1} & L_{2} & L \\ 
M_{1} & M_{2} & M
\end{array}
\right) \left\{ 
\begin{array}{ccc}
L_{1} & L_{2} & L \\ 
j & j & j
\end{array}
\right\}  \nonumber \\
&&\qquad \times \sqrt{(2L_{1}+1)(2L_{2}+1)(2L+1)}.  \label{B}
\end{eqnarray}
In terms of the irreducible tensors, the operator $|jm\rangle \langle
jm^{\prime }|$ is expressed as follows: 
\begin{equation}
|jm\rangle \langle jm^{\prime
}|=\sum_{L=0}^{2j}\sum_{M=-L}^{L}(-1)^{j-m^{\prime }}\langle jm;j-m^{\prime
}|LM\rangle \,\hat{T}_{LM}^{(j)}\ .  \label{equation16}
\end{equation}
This means that the irreducible tensors are a basis for the linear space of
operators acting on the Hilbert space of the $SU(2)$ irreducible
representation.

\subsection{Tomogram spin symbol and reconstruction formula}

The tomogram symbol of the observable $\hat{A}^{(j)}$ is 
\begin{eqnarray}
w\left( m_{1},\beta ,\gamma \right) &=&\mbox {Tr}\left[ \hat{A}%
^{(j)}R^{\dagger }(g)\,|jm_{1}\rangle \langle jm_{1}|\,R(g)\right]  \nonumber
\\
&=&\sum_{m_{1}^{\prime }=-j}^{j}\,\sum_{m_{2}^{\prime
}=-j}^{j}\,D_{m_{1}m_{1}^{\prime }}^{(j)}(\alpha ,\beta ,\gamma
)\,A_{m_{1}^{\prime }m_{2}^{\prime }}^{(j)}\,D_{m_{1}m_{2}^{\prime
}}^{(j)*}(\alpha ,\beta ,\gamma )\,.  \label{equation05}
\end{eqnarray}
In view of (\ref{equation05}), the tomogram depends only on two Euler
angles, i.e., the tomogram depends on the spin projection and on a point on
the Bloch sphere.

The tomogram can be presented in another form using a Kronecker
delta-function, which is the general form for tomograms of arbitrary
observables suggested in \cite{Manko1} 
\begin{equation}
w(m_{1},\beta ,\gamma )=\mbox{Tr}\,\hat{A}^{(j)}\delta \Big( %
m_{1}-R^{\dagger }(g)\hat{J}_{3}R(g)\Big).  \label{new1}
\end{equation}
It is obvious that the tomogram of the identity operator is the unit.

To derive the inverse of~(\ref{equation05}), we multiply by the Wigner $D$%
-function $D_{\mu ^{\prime }m^{\prime }}^{j^{\prime }}(\alpha ,\beta ,\gamma
)$ and integrate over the volume element of the $SU(2)$ group, i.e., 
\begin{eqnarray}
&&\int d\Omega \,\,w(m_{1},\beta ,\gamma )\,D_{\mu ^{\prime }m^{\prime
}}^{j^{\prime }}(\alpha ,\beta ,\gamma )  \nonumber \\
&=&\sum_{m_{1}^{\prime }m_{2}^{\prime }}\langle j^{\prime }m^{\prime
};jm_{1}^{\prime }|jm_{2}^{\prime }\rangle \langle j^{\prime }\mu ^{\prime
};jm_{1}|jm_{1}\rangle \,\frac{8\pi ^{2}}{2j+1}\,A_{m_{1}^{\prime
}m_{2}^{\prime }}^{j},  \label{equation19}
\end{eqnarray}
where the known property of the Wigner $D$-functions $\Big(D(\alpha ,\beta
,\gamma )\equiv D(\Omega )\Big)$ 
\begin{eqnarray}
&&\int d\Omega \,\,D_{m_{3}^{\prime }m_{3}}^{j_{3}\,*}(\Omega
)\,D_{m_{2}^{\prime }m_{2}}^{j_{2}}(\Omega )\,D_{m_{1}^{\prime
}m_{1}}^{j_{1}}(\Omega )  \nonumber \\
&=&\frac{8\pi ^{2}}{2j+1}\langle j_{1}m_{1};j_{2}m_{2}|j_{3}m_{3}\rangle
\langle j_{1}m_{1}^{\prime };j_{2}m_{2}^{\prime }|j_{3}m_{3}^{\prime
}\rangle.  \label{equation20}
\end{eqnarray}
was used.

In view of the symmetry relations and properties of the Clebsch--Gordan
coefficients, we have that
\begin{eqnarray}
&&\langle j^{\prime }m^{\prime };jm_{1}^{\prime }|jm_{2}^{\prime }\rangle
\langle j^{\prime }\mu ^{\prime };jm_{1}|jm_{2}\rangle   \nonumber \\
&=&(-1)^{j+m_{1}+j+m_{1}^{\prime }}\,\frac{2j+1}{2j^{\prime }+1}\delta _{\mu
^{\prime }\,0}\,\langle jm_{1};j-m_{1}|j^{\prime }0\rangle \langle
jm_{2}^{\prime };j-m_{1}^{\prime }|j^{\prime }m^{\prime }\rangle .
\label{equation21}
\end{eqnarray}
Using the orthonormality property of Clebsch--Gordan coefficients 
\[
\sum_{m_{1}=-j}^{j}\langle jm_{1};j-m_{1}|j^{\prime }0\rangle \langle
jm_{1};j-m_{1}|j^{\prime }0\rangle =1,
\]
we have
\begin{eqnarray}
&&\sum_{m_{1}}\frac{2j^{\prime }+1}{8\pi ^{2}}\langle
jm_{1};j-m_{1}|j^{\prime }0\rangle \,\int d\Omega
\,\,(-1)^{j+m_{1}}\,w(m_{1},\beta ,\gamma )\,D_{0m^{\prime }}^{j^{\prime
}}(\Omega )  \nonumber \\
&=&\sum_{m_{1}^{\prime },m_{2}^{\prime }=-j}^{j}(-1)^{j+m_{1}^{\prime
}}\langle jm_{2}^{\prime };j-m_{1}^{\prime }|j^{\prime }m^{\prime }\rangle
A_{m_{1}^{\prime }m_{2}^{\prime }}^{(j)}\ .  \label{equation22}
\end{eqnarray}
Multiplying this equation by $\langle j\mu _{2};j\mu _{1}|j^{\prime
}m^{\prime }\rangle $ and summing over the indexes $j^{\prime }$ and $%
m^{\prime }$ we arrive at the result
\begin{eqnarray}
&&A_{\mu _{1}\mu _{2}}^{(j)}=\sum_{j^{\prime }=0}^{2j}\sum_{m^{\prime
}=-j^{\prime }}^{j^{\prime }}\sum_{m_{1}=-j}^{j}(-1)^{m_{1}-\mu
_{1}}\,\langle jm_{1};j-m_{1}|j^{\prime }0\rangle \langle j\mu _{1};j-\mu
_{2}|j^{\prime }\mu ^{\prime }\rangle   \nonumber \\
&&\qquad \times \int d\Omega \,w(m_{1},\beta ,\gamma )\,D_{0-m^{\prime
}}^{j^{\prime }}(\alpha ,\beta ,\gamma )\ .  \label{equation23}
\end{eqnarray}
Using Eqs. (\ref{eq.2}) and (\ref{equation16}) we can write the observable
operator $\hat{A}^{(j)}$ in terms of unitary irreducible tensors as follows: 
\begin{equation}
\hat{A}^{(j)}=\sum_{\mu _{1},\mu
_{2}=-j}^{j}\sum_{L=0}^{2j}\sum_{M=-L}^{L}(-1)^{j-\mu _{2}}\langle j\mu
_{1};j-\mu _{2}|LM\rangle \,\hat{T}_{LM}^{(j)}\,A_{\mu _{1}\mu _{2}}^{(j)}\ .
\label{equation24}
\end{equation}
Substituting $A_{\mu _{1}\mu _{2}}^{(j)}$ into (\ref{equation24}), in view
of the orthonormality of the Clebsch--Gordan coefficients, we obtain the
observable in terms of its tomogram: 
\begin{eqnarray}
&&\hat{A}^{(j)}=\sum_{L=0}^{2j}\sum_{M=-L}^{L}\sum_{m=-j}^{j}(-1)^{j-m+M}%
\frac{2L+1}{8\pi ^{2}}\langle jm;j-m|L0\rangle   \nonumber \\
&&\qquad \times \left( \int d\Omega \,w(m,\beta ,\gamma
)\,D_{0-M}^{L}(\alpha ,\beta ,\gamma )\right) \,\hat{T}_{LM}^{(j)}.
\label{equation25}
\end{eqnarray}

The density operator $\hat{\rho}$ can be expanded in terms of irreducible
tensors (\ref{TLM}) as follows: 
\begin{eqnarray}
\hat{\rho} &=&\sum_{L=0}^{2j}\sum_{M=-L}^{L}(-1)^{M}\,\frac{2L+1}{8\pi ^{2}}%
\int d\Omega \,D_{0-M}^{(L)}(\alpha ,\beta ,\gamma )  \nonumber \\
&&\times \sum_{m=-j}^{j}\,(-1)^{j-m}\,w(m,\beta ,\gamma )\,\langle
jm;j-m|L0\rangle \,\hat{T}_{LM}^{(j)}\ .  \label{equation29}
\end{eqnarray}

One can express the operators determining the star-product of tomographic
symbols in terms of irreducible tensors. By comparing the formulas defining
the generic symbol of operators (\ref{eq.A1}) and its inverse (\ref{eq.A2})
with the formulae defining the observable tomogram (\ref{new1}) and its
inverse (\ref{equation25}), one can find the operators $\hat{U}(\mathbf{x})$
and $\hat{D}(\mathbf{x})$ explicitly. The operators $\hat{U}(\mathbf{x}%
)\equiv \hat{U}(m,\Omega )$ and $\hat{D}(\mathbf{x}\equiv \hat{D}(m,\Omega )$
can be expressed as follows: 
\begin{eqnarray}
&&\hat{U}(m,\Omega )=\sum_{L=0}^{2j}\,\sum_{M=-L}^{L}(-1)^{j-m+M}\,\langle
jm;j-m|L0\rangle \,\,D_{0-M}^{L}(\alpha ,\beta ,\gamma )\,\hat{T}_{LM}^{(j)},
\label{neweq37} \\
&&\hat{D}(m,\Omega )=\sum_{L=0}^{2j}\,\sum_{M=-L}^{L}(-1)^{j-m+M}\frac{2L+1}{%
8\pi ^{2}}\,\langle jm;j-m|L0\rangle \,\,D_{0-M}^{L}(\alpha ,\beta ,\gamma
)\,\hat{T}_{LM}^{(j)}.  \nonumber \\
&&  \label{neweq38}
\end{eqnarray}

\subsection{The kernel of the star-product}

Using formulae~(\ref{neweq37}) and (\ref{neweq38}), one can write down a
composition rule for two symbols $f_{\hat{A}}(\mathbf{x})$ and $f_{\hat{B}}(%
\mathbf{x})$ determining the star-product of these symbols. The composition
rule is 
\begin{equation}
f_{\hat{A}}(\mathbf{x})*f_{\hat{B}}(\mathbf{x})=\int f_{\hat{A}}(\mathbf{x}%
^{\prime \prime })f_{\hat{B}}(\mathbf{x}^{\prime })K(\mathbf{x}^{\prime
\prime },\mathbf{x}^{\prime },\mathbf{x})\,d\mathbf{x}^{\prime }\,d\mathbf{x}%
^{\prime \prime }.  \label{eq.25}
\end{equation}
The kernel in the integral of (\ref{eq.25}) is the trace of the product of
the operators used to construct the map 
\begin{equation}
K(\mathbf{x}^{\prime \prime },\mathbf{x}^{\prime },\mathbf{x})=\mbox{Tr}%
\left[ \hat{D}(\mathbf{x}^{\prime \prime })\hat{D}(\mathbf{x}^{\prime })\hat{%
U}(\mathbf{x})\right] .  \label{eq.26}
\end{equation}
Within this framework, according to (\ref{a01}), (\ref{a02}) and (\ref{a04}%
), one has two equivalent expressions for the operator $\hat{U}(\mathbf{x})$ 
\begin{equation}
\hat{U}(\mathbf{x})=\delta \Big(m_{1}-R^{\dagger }(g)\hat{J}_{3}R(g)\Big) %
=R(g)^{\dagger }\mid jm_{1}\rangle \langle jm_{1}\mid R(g)  \label{new2}
\end{equation}
or, due to the structure of this equation, 
\begin{equation}
\hat{U}(\mathbf{x})=\delta (m_{1}-\mathbf{n}\cdot \hat{\mathbf{J}}),\qquad 
\mathbf{n}=\left( \sin \beta \cos \gamma ,\sin \beta \sin \gamma ,\cos \beta
\right) .  \label{new3}
\end{equation}
The dual operator reads 
\begin{equation}
\hat{D}(\mathbf{x})=\sum_{L=0}^{2j}\sum_{M=-L}^{L}(-1)^{j-m+M}\,\frac{2L+1}{%
8\pi ^{2}}D_{0-M}^{(L)}(\alpha ,\beta ,\gamma )\langle jm;j-m|L0\rangle \,%
\hat{T}_{LM}^{(j)},  \label{new4}
\end{equation}
where $\hat{T}_{LM}^{(j)}$ is given in Eq.~(\ref{TLM}).

Inserting the expressions for the operators $\hat{U}(\mathbf{x})$ and $\hat{D%
}(\mathbf{x})$ in (\ref{eq.26}) and using the properties of irreducible
tensors (\ref{A}) and (\ref{B}), one obtains an explicit form for the kernel
of the spin star-product
\begin{eqnarray}
&&K(\mathbf{x}_{2},\mathbf{x}_{1},\mathbf{x})\equiv K(m_{2},\Omega
_{2},m_{1},\Omega _{1},m,\Omega )\phantom{flush this to the left flush this
to the left}  \nonumber \\
&=&(-1)^{j-m-m_{1}-m_{2}}\sum_{L=0}^{2j}\sum_{L_{1}=0}^{2j}%
\sum_{L_{2}=0}^{2j}\,\frac{(2L_{1}+1)(2L_{2}+1)}{64\pi ^{4}}  \nonumber \\
&&\ \times \langle jm;j-m|L0\rangle \,\langle jm_{1};j-m_{1}|L_{1}0\rangle
\,\langle jm_{2};j-m_{2}|L_{2}0\rangle   \nonumber \\
&&\ \times
\sum_{M=-L}^{L}\sum_{M_{1}=-L_{1}}^{L_{1}}%
\sum_{M_{2}=-L_{2}}^{L_{2}}(-1)^{L+L_{1}+L_{2}}\sqrt{%
(2L+1)(2L_{1}+1)(2L_{2}+1)}  \nonumber \\
&&\ \times \left\{ 
\begin{array}{ccc}
L_{1} & L_{2} & L \\ 
j & j & j
\end{array}
\right\} \,\left( 
\begin{array}{ccc}
L_{1} & L_{2} & L \\ 
M_{1} & M_{2} & M
\end{array}
\right) \,D_{0-M}^{(L)}(\Omega )\,D_{0-M_{1}}^{(L_{1})}(\Omega
_{1})\,D_{0-M_{2}}^{(L_{2})}(\Omega _{2})\,.  \nonumber \\
&&  \label{eq.26a}
\end{eqnarray}

\subsection{Unitary spin tomography}

One can extend the construction by introducing a unitary spin tomogram \cite
{SudPLA} of the multiqudit state with density matrix $\rho$. For this, one
uses the joint probability distribution 
\begin{equation}
w(m_{1},m_{2},\ldots ,m_{M},u)=\langle m_{1},m_{2},\ldots ,m_{M}\mid
u^{\dagger }\rho u\mid m_{1},m_{2},\ldots ,m_{M}\rangle ,  \label{A86}
\end{equation}
where $u$ is a unitary operator in the Hilbert space of multiqudit states.

For a simple qudit state, the tomogram unitary symbol is 
\begin{equation}
w(m_{1},u_{1})=\langle m_{1}\mid u_{1}^{\dagger }\rho u_{1}\mid m_{1}\rangle
,  \label{A87}
\end{equation}
where $u_{1}$ is a $(2j+1)\times (2j+1)$ matrix.

Since it is possible to reconstruct the density matrix using only spin
tomograms, the unitary spin tomogram also determines the density matrix
completely. One can integrate in Eq. (\ref{equation29}) the unitary spin
tomogram $w(m,u)$ using the Haar measure $\stackrel{\symbol{126}}{du}$
instead of $d\Omega $ and adding the delta-function term $\delta \Big(%
u-D_{0-M}^{(L)}(\alpha ,\beta ,\gamma )\Big)$. This construction means that
the spin quantum state is defined by a map of the unitary group to the
simplex.

The following are the properties of the unitary spin tomograms of multiqudit
systems:

(i)Normalization 
\begin{equation}  \label{GG1}
\sum_{\vec m}w(\vec m,u)=1,\qquad w(\vec m,u)\geq 0;
\end{equation}

(ii)Group normalization

From the Haar measure on the unitary group $\tilde {d}u$ divided by the
group volume $V=\int \tilde{d}u$, one obtains the measure $du=\tilde{d}u/V$
with $\int du=1$. Then, 
\begin{equation}
\int du\,w(\vec{m},u)=1.  \label{GG2}
\end{equation}
This property follows from the orthogonality condition for matrix elements
of unitary matrices as elements of an irreducible representation of a
compact group. Another property is 
\begin{equation}
\sum_{\vec{m}_{2}}w(\vec{m},\vec{m}_{2},u_{1}\otimes u_{2})=w(\vec{m}%
_{1},u_{1}),\quad \vec{m}=(\vec{m}_{1},\vec{m}_{2}),  \label{GG3}
\end{equation}
where the tomogram $w(\vec{m}_{1},u_{1})$ is a tomogram for the subsystem
density matrix $\hat{\rho}_{1}=\mbox{Tr}_{2}\,\hat{\rho}_{12}$.

An analogous unitary group integration property follows from the relation 
\begin{equation}
\int u_{js}A_{\ldots sm\ldots }u_{mk}^{\dagger }\,du=\delta _{jk}A_{\ldots
ss\ldots },  \label{GG4}
\end{equation}
yielding
\begin{equation}
\int w(\vec{m}_{1},\vec{m}_{2},u_{1}\times u_{2})\,du_{2}=w(\vec{m}%
_{1},u_{1}),  \label{GG5}
\end{equation}
that corresponds to 
\begin{equation}
\int u_{2}^{\dagger }\rho _{12}u_{2}\,du_{2}=\mbox{Tr}_{2}\,\rho _{12}.
\label{GG6}
\end{equation}

\section{Operator symbols as maps from the unitary group to the simplex}

The unitary spin symbol (\ref{A86}) defines, for each density matrix $\rho $%
, a mapping from the unitary group $U\left( N\right) $, $N=%
\prod_{k=1}^{M}(2j_{k}+1)$, to a $N-1$ dimensional simplex. The nature of
the image of $U\left( N\right) $ on the simplex depends on the nature of the
density matrix.

\begin{teo}
The unitary spin symbol image of $U\left( N\right) $ on the simplex for most
density matrices $\rho $ ($\rho ^{\prime }$s with at least two different
eigenvalues) has dimension $N-1$. For pure states, it is the whole simplex
and for mixed states a volume bounded by the hyperplanes
\end{teo}

\begin{equation}
\begin{array}{ccc}
\lambda _{\min }\leq x_{i}\leq \lambda _{\max }, &  & i=1,\cdots ,N-1, \\ 
\lambda _{\min }\leq \left( 1-\sum_{i=1}^{N-1}x_{i}\right) \leq \lambda
_{\max }, &  & 
\end{array}
\label{M1}
\end{equation}
where $\left\{ \lambda _{k}\right\} $ are the eigenvalues of the density
matrix.

\textit{Proof}: $\exists u$ such that $u^{\dagger }\rho u=\left( \lambda
_{1},\lambda _{2},...\lambda _{N}\right) $ is diagonal. Then by another $%
u^{\prime }$ 
\begin{equation}
w(m_{1},m_{2},\ldots ,m_{N},uu^{\prime })=\left\{ \sum_{k}\left|
u_{kj}^{\prime }\right| ^{2}\lambda _{k},j=1,...,N\right\}.  \label{M2}
\end{equation}
If $\rho $ is a pure state, only one $\lambda _{i}\neq 0$. Then 
\[
w(m_{1},m_{2},\ldots ,m_{N},uu^{\prime })=\left\{ \left| u_{1j}^{\prime
}\right| ^{2},j=1,...,n\right\}, 
\]
that is, all points in the simplex are obtained. Therefore, for a pure
state, the unitary tomographic symbol maps the unitary group on the whole
simplex.

To obtain the dimensionality of the image for a general (mixed) state, we
consider the elementary $U\left( N\right) $ transformations : 
\begin{equation}
\begin{array}{lll}
d_{k}\left( \varphi \right) =\mathnormal{diag}\left( 1,,1,e^{i\varphi
},1,1\right)  &  &  \\ 
g_{ij}\left( \theta \right) =\left( 
\begin{array}{ll}
\cos \theta  & \sin \theta  \\ 
-\sin \theta  & \cos \theta 
\end{array}
\right)  &  & \mbox{in\,\,\,coordinates}\,\,ij \\ 
g_{ij}^{C}\left( \theta \right) =\left( 
\begin{array}{ll}
\cos \theta  & i\sin \theta  \\ 
i\sin \theta  & \cos \theta 
\end{array}
\right)  &  & \mbox{in\,\,\, coordinates}\,\,ij
\end{array}
\label{M3}
\end{equation}
Consider these elementary transformations acting on the diagonalized matrix $%
u^{\dagger }\rho u$. $d_{k}\left( \varphi \right) $ does not change the
diagonal elements and both $g_{ij}\left( \theta \right) $ and $%
g_{ij}^{C}\left( \theta \right) $ have a similar action: 
\begin{equation}
\begin{array}{lll}
\lambda _{i} & \rightarrow  & \lambda _{i}\cos ^{2}\theta +\lambda _{j}\sin
^{2}\theta , \\ 
\lambda _{j} & \rightarrow  & \lambda _{i}\sin ^{2}\theta +\lambda _{j}\cos
^{2}\theta .
\end{array}
\label{M4}
\end{equation}
A general infinitesimal transformation would be 
\[
\begin{array}{lll}
\lambda _{i} & \rightarrow  & \lambda _{i}+\sum_{k\neq i}\alpha _{ik}\left(
\lambda _{k}-\lambda _{i}\right) \hspace{2cm}\left( \alpha _{ik}=\alpha
_{ki}\right) 
\end{array}
\]
and the dimension of the simplex image of $U\left( N\right) $ is the rank of
the Jacobian $\frac{\partial \lambda }{\partial \alpha }$. If $\rho $ has at
least two different eigenvalues, the rank is $N-1$, this being the dimension
of the simplex image. The hyperplanes (\ref{M1}) bounding this simplex
volume follow from the convex nature of the eigenvalues linear combination (%
\ref{M2}). The situation where all eigenvalues are equal is exceptional, the
image being a point in this case. {\raggedright $\Box $}

Figure 1 shows an example for a mixed state of a two-qubit state, when $%
\lambda _{1}=0.4,\lambda _{2}=0.3,\lambda _{3}=0.2,\lambda _{4}=0.1$.

\begin{figure}[htb]
\begin{center}
\psfig{figure=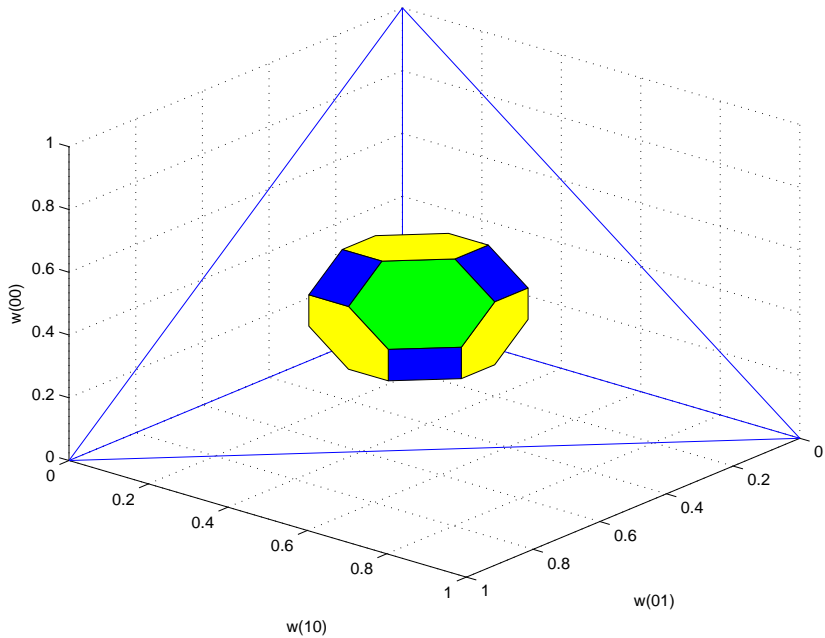,width=11truecm}
\end{center}
\caption{Simplex image of a mixed two qubit state $\lambda _{1}=0.4,\lambda
_{2}=0.3,\lambda _{3}=0.2,\lambda _{4}=0.1$.}
\end{figure}

For a bipartite system of dimension $N_{1}\times N_{2}$, the distinction
between factorized and entangled states refers to the behavior under
transformations of the factorized group $U(N_{1})\otimes 1+1\otimes U(N_{2})$%
. We call a state \textit{factorized}, if the density matrix is 
\[
\rho =\rho ^{(1)}\otimes \rho ^{(2)}, 
\]
and \textit{classically correlated}, if 
\[
\rho =\sum_{k=1}^{n}c_{k}\rho _{k}^{(1)}\otimes \rho _{k}^{(2)}, 
\]
with $\sum_{k=1}^{n}c_{k}=1$.

\begin{teo}
The simplex symbol image under $G_{12}=U(N_{1})\otimes 1+1\otimes U(N_{2})$
of a generic factorized or classically correlated state has dimension $%
\left( N_{1}-1\right) +\left( N_{2}-1\right) $.
\end{teo}

\textit{Proof} : For a classically correlated state, if $n>1$ it is not, in
general, possible to find an element of $G_{12}$ diagonalizing $\rho $.
Therefore, one has to consider the action of the elementary unitary
transformations (\ref{M3}) on a general matrix. $d_{k}\left( \varphi \right) 
$ does not change the diagonal elements whereas the $g_{ij}\left( \theta
\right) $ action for $1\otimes U(N_{2})$ is 
\[
\begin{array}{lll}
\sum_{k=1}^{n}c_{k}\left( \rho _{k}^{(1)}\right) _{aa}\otimes \left( \rho
_{k}^{(2)}\right) _{ii} & \rightarrow & 
\begin{array}{l}
\sum_{k=1}^{n}c_{k}\left( \rho _{k}^{(1)}\right) _{aa}\otimes \left\{ \left(
\rho _{k}^{(2)}\right) _{ii}\cos ^{2}\theta \right. \\ 
\left. +\left( \rho _{k}^{(2)}\right) _{jj}\sin ^{2}\theta -2\mbox{Re}\left(
\rho _{k}^{(2)}\right) _{ij}\sin \theta \cos \theta \right\},
\end{array}
\\ 
\sum_{k=1}^{n}c_{k}\left( \rho _{k}^{(1)}\right) _{aa}\otimes \left( \rho
_{k}^{(2)}\right) _{jj} & \rightarrow & 
\begin{array}{l}
\sum_{k=1}^{n}c_{k}\left( \rho _{k}^{(1)}\right) _{aa}\otimes \left\{ \left(
\rho _{k}^{(2)}\right) _{ii}\sin ^{2}\theta \right. \\ 
\left. +\left( \rho _{k}^{(2)}\right) _{jj}\cos ^{2}\theta +2\mbox{Re}\left(
\rho _{k}^{(2)}\right) _{ij}\sin \theta \cos \theta \right\},
\end{array}
\end{array}
\]
and for $g_{ij}^{C}\left( \theta \right) $ is 
\[
\begin{array}{lll}
\sum_{k=1}^{n}c_{k}\left( \rho _{k}^{(1)}\right) _{aa}\otimes \left( \rho
_{k}^{(2)}\right) _{ii} & \rightarrow & 
\begin{array}{l}
\sum_{k=1}^{n}c_{k}\left( \rho _{k}^{(1)}\right) _{aa}\otimes \left\{ \left(
\rho _{k}^{(2)}\right) _{ii}\cos ^{2}\theta \right. \\ 
\left. +\left( \rho _{k}^{(2)}\right) _{jj}\sin ^{2}\theta -2\mbox{Im}\left(
\rho _{k}^{(2)}\right) _{ij}\sin \theta \cos \theta \right\},
\end{array}
\\ 
\sum_{k=1}^{n}c_{k}\left( \rho _{k}^{(1)}\right) _{aa}\otimes \left( \rho
_{k}^{(2)}\right) _{jj} & \rightarrow & 
\begin{array}{l}
\sum_{k=1}^{n}c_{k}\left( \rho _{k}^{(1)}\right) _{aa}\otimes \left\{ \left(
\rho _{k}^{(2)}\right) _{ii}\sin ^{2}\theta \right. \\ 
\left. +\left( \rho _{k}^{(2)}\right) _{jj}\cos ^{2}\theta +2\mbox{Im}\left(
\rho _{k}^{(2)}\right) _{ij}\sin \theta \cos \theta \right\}.
\end{array}
\end{array}
\]
For generic $\rho $ matrices, $U(N_{1})\otimes 1$ and $1\otimes U(N_{2})$
operate independently, therefore, infinitesimal transformations explore $%
\left( N_{1}-1\right) +\left( N_{2}-1\right) $ independent directions. {%
\raggedright $\Box $}

The generalization to classically correlated multipartite systems is
immediate, implying that the image dimension under $\sum_{i}1\otimes \cdots
\otimes U(N_{i})\otimes \cdots \otimes 1$ is $\sum_{i}\left( N_{i}-1\right) $%
.

As an example, we compute explicitly the equation for the two-dimensional
surface image in the two-qubit case for a factorized state. In this case,
one has to consider mappings from $U(2)\otimes 1+1\otimes U(2)$ to the
simplex.

Let $\rho =\rho _{1}\otimes \rho _{2}$. Here, without loosing generality, $%
\rho $ may be considered as diagonal. Then 
\[
\Big( u(a)\otimes 1+1\otimes u^{\prime }(b)\Big) ^{\dagger }\left( \rho
_{1}\otimes \rho _{2}\right) \Big( u(a)\otimes 1+1\otimes u^{\prime }(b)%
\Big)
\]
is 
\begin{eqnarray*}
&&\left( 
\begin{array}{cc}
\left| a_{11}\right| ^{2}\lambda _{1}+\left| a_{21}\right| ^{2}\lambda _{2}
& \ldots \\ 
\ldots & \left| a_{12}\right| ^{2}\lambda _{1}+\left| a_{22}\right|
^{2}\lambda _{2}
\end{array}
\right) \\
&&\otimes \left( 
\begin{array}{cc}
\left| b_{11}\right| ^{2}\mu _{1}+\left| b_{21}\right| ^{2}\mu _{2} & \ldots
\\ 
\ldots & \left| b_{12}\right| ^{2}\mu _{1}+\left| b_{22}\right| ^{2}\mu _{2}
\end{array}
\right).
\end{eqnarray*}
Hence 
\[
w\left( m_{1},m_{2};u(a)\otimes 1+1\otimes u^{\prime }(b)\right) 
\]
is 
\[
\begin{array}{lll}
w\left( 00\right) & = & \left( \lambda _{1}\cos ^{2}\frac{\theta }{2}%
+\lambda _{2}\sin ^{2}\frac{\theta }{2}\right) \left( \mu _{1}\cos ^{2}\frac{%
\alpha }{2}+\mu _{2}\sin ^{2}\frac{\alpha }{2}\right), \\ 
w\left( 01\right) & = & \left( \lambda _{1}\cos ^{2}\frac{\theta }{2}%
+\lambda _{2}\sin ^{2}\frac{\theta }{2}\right) \left( \mu _{1}\sin ^{2}\frac{%
\alpha }{2}+\mu _{2}\cos ^{2}\frac{\alpha }{2}\right), \\ 
w\left( 10\right) & = & \left( \lambda _{1}\sin ^{2}\frac{\theta }{2}%
+\lambda _{2}\cos ^{2}\frac{\theta }{2}\right) \left( \mu _{1}\cos ^{2}\frac{%
\alpha }{2}+\mu _{2}\sin ^{2}\frac{\alpha }{2}\right), \\ 
w\left( 11\right) & = & \left( \lambda _{1}\sin ^{2}\frac{\theta }{2}%
+\lambda _{2}\cos ^{2}\frac{\theta }{2}\right) \left( \mu _{1}\sin ^{2}\frac{%
\alpha }{2}+\mu _{2}\cos ^{2}\frac{\alpha }{2}\right),
\end{array}
\]
implying 
\[
w\left( 10\right) =\frac{w\left( 00\right) }{w\left( 01\right) +w\left(
00\right) }-w\left( 00\right). 
\]
Figure 2 shows this two-dimensional surface in the 3-dimensional simplex.

For a pure state, this would be the image of the $U(2)\otimes 1+1\otimes U(2)
$ group.

For a mixed state, the image is the intersection of the surface with the
spanned volume, as in Fig. 1.

\begin{figure}[htb]
\begin{center}
\psfig{figure=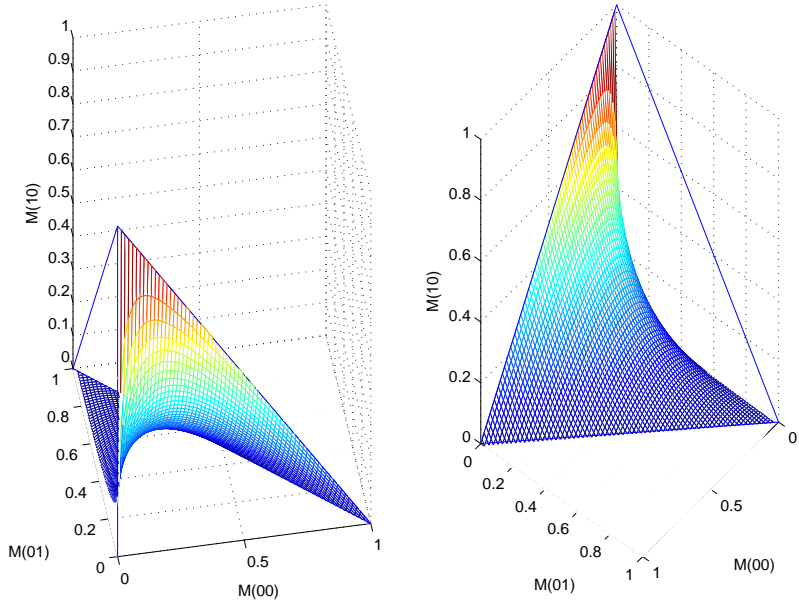,width=11truecm}
\end{center}
\caption{The two-dimensional surface image of $U(2)\otimes 1+1\otimes U(2)$
in the two-qubit case for a factorized pure state.}
\end{figure}

Theorem 4.2 suggests a notion of \textit{geometric correlation}, namely,

\begin{defin}
A state of a multipartite system is called geometrically correlated if the
symbol image under $\sum_{i}1\otimes \cdots \otimes U(N_{i})\otimes \cdots
\otimes 1$ has dimension less than $\sum_{i}\left( N_{i}-1\right) $.
\end{defin}

Deviations from geometrical genericity occur when the systems are entangled
or the density matrix has special symmetry properties.

As an example, consider the entangled state 
\[
c_{0}\mid 00\rangle +c1\mid 11\rangle. 
\]
A simple computation shows that the image under $U(2)\otimes 1+1\otimes U(2)$
is defined by 
\[
\begin{array}{lll}
\frac{w\left( 00\right) }{\left| c_{0}\right| ^{2}} & = & \frac{w\left(
11\right) }{\left| c1\right| ^{2}}, \\ 
\frac{w\left( 10\right) }{c_{1}c_{0}^{*}} & = & \frac{w\left( 01\right) }{%
c_{0}c_{1}^{*}},
\end{array}
\]
implying that the image is one-dimensional (Figure 3)

\begin{figure}[tbh]
\begin{center}
\psfig{figure=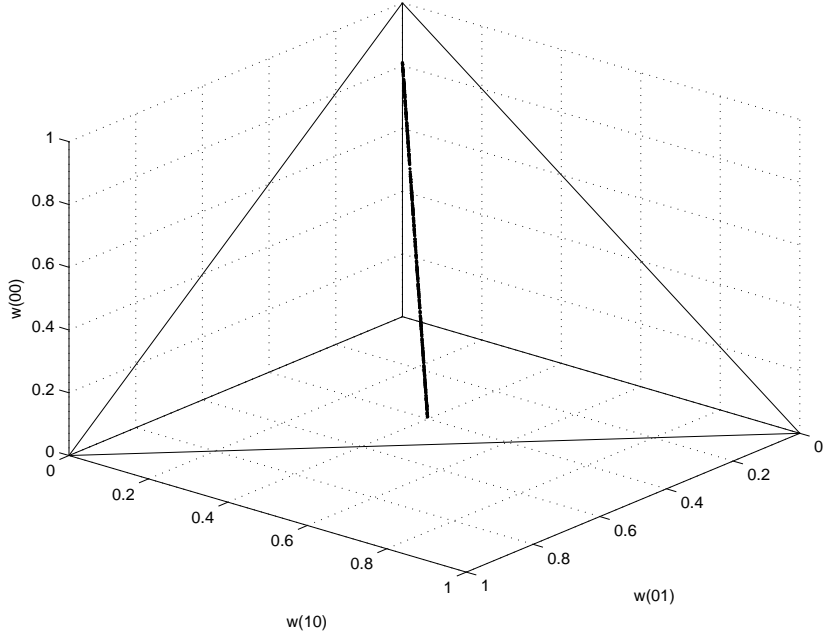,width=11truecm}
\end{center}
\caption{The simplex image of an entangled state.}
\end{figure}

However, the dimension reduction of the image of $U(2)\otimes 1+1\otimes
U(2) $ does not coincide with the notion of entanglement. As an example,
consider the Werner state 
\[
\rho _{W}=\frac{1}{4}\left( 
\begin{array}{llll}
1-q & 0 & 0 & 0 \\ 
0 & 1+q & -2q & 0 \\ 
0 & -2q & 1+q & 0 \\ 
0 & 0 & 0 & 1-q
\end{array}
\right), 
\]
$q\leq 1$, which is known to be entangled only for $q>\frac{1}{3}$. In this
case, because of the highly symmetric nature of the state, the orbit of the
tomgraphic symbol is the same both for $U(2)\otimes 1$ and $1\otimes U(2)$,
namely, 
\[
w=\frac{1}{4}\left( 
\begin{array}{l}
\left( 1-q\right) \cos ^{2}\theta +\left( 1+q\right) \sin ^{2}\theta \\ 
\left( 1+q\right) \cos ^{2}\theta +\left( 1-q\right) \sin ^{2}\theta \\ 
\left( 1+q\right) \cos ^{2}\theta +\left( 1-q\right) \sin ^{2}\theta \\ 
\left( 1-q\right) \cos ^{2}\theta +\left( 1+q\right) \sin ^{2}\theta
\end{array}
\right), 
\]
implying that the image is always one-dimensional.

Incidentally, Peres separability criterium \cite{Peres} applied to the
partial transpose 
\[
\rho _{W}^{t_{B}}=\frac{1}{4}\left( 
\begin{array}{llll}
1-q & 0 & 0 & -2q \\ 
0 & 1+q & 0 & 0 \\ 
0 & 0 & 1+q & 0 \\ 
-2q & 0 & 0 & 1-q
\end{array}
\right) 
\]
expressed in tomographic operator symbols would be 
\[
\sum_{\left\{ m_{i}\right\} }\left| w_{\rho _{W}^{t_{B}}}(\left\{
m_{i}\right\} ,u)\right| =1\hspace{2cm}\forall u, 
\]
the state being entangled when there is an $u$ for which this identity is
violated.

\section{Measurements and generalized measurements}

In the standard quantum formulation, measurements are realized by von
Neumann ``instruments'' which are orthogonal projectors $\hat{P}_{j}$ onto
eigenstates of the variables being measured. The projectors applied to the
pure state $\mid \psi \rangle $ yield 
\begin{equation}
\mid \psi \rangle _{j}=\hat{P}_{j}\mid \psi \rangle ,\quad \hat{P}_{j}=\mid
\psi _{j}\rangle \langle \psi _{j}\mid ,  \label{AI1A}
\end{equation}
or in terms of density matrix $\mid \psi \rangle \langle \psi \mid $, one
has the result 
\begin{equation}
\mid \psi \rangle _{j}\,_{j}\langle \psi \mid =\hat{P}_{j}\mid \psi \rangle
\langle \psi \mid \hat{P}_{j}=\hat{P}_{j}\hat{\rho}_{\psi }\hat{P}%
_{j}=|\langle \psi \mid \psi _{j}\rangle |^{2}\mid \psi _{j}\rangle \langle
\psi _{j}\mid .  \label{AI2A}
\end{equation}

For a mixed state $\hat{\rho}$, the measurement provides the state density
operator after measurement 
\begin{equation}
\hat{\rho}_{j}=\hat{P}_{j}\hat{\rho}\hat{P}_{j}.  \label{AI4A}
\end{equation}

Generalized measurements use positive operator-valued measures (POVM), that
is, positive operators $\hat{P}_{k}$ with the property 
\begin{equation}
\sum_{k}\hat{P}_{k}=\hat{1},  \label{AI5A}
\end{equation}
the index $k$ being either discrete or continuous. In the latter case, one
has an integration in (\ref{AI5A}).

Within the framework of operator symbols and star-products, instead of (\ref
{AI4A}), we have after the measurement a symbol for the density operator of
the state 
\begin{equation}
f_{\hat{\rho}_{j}}(\mathbf{x})=f_{\hat{P}_{j}}(\mathbf{x})\star f_{\hat{\rho}%
}(\mathbf{x})\star f_{\hat{P}_{j}}(\mathbf{x}),  \label{AI6A}
\end{equation}
$f_{\rho }(\mathbf{x})$ being the symbol of the density operator of
measurable state and $f_{\hat{P}_{j}}(\mathbf{x})$ the symbol of the
instrument.

In the tomographic probability representation, the result of measurements is
described by a map of the probability distributions, namely, 
\begin{equation}
w_{j}(X,\mu ,\nu )=w_{\hat{P}_{j}}(X,\mu ,\nu )\star w(X,\mu ,\nu )\star w_{%
\hat{P}_{j}}(X,\mu ,\nu ),  \label{AI7A}
\end{equation}
with the kernel of the star-product of tomograms given by Eq. (\ref{KERNEL}).

For the case of spin (or unitary spin) tomograms, the linear map of
tomographic-probability distributions is realized by the formula 
\begin{equation}
w_{j}(m,\vec{n})=w_{\hat{P}_{j}}(m,\vec{n})\star w(m,\vec{n})\star w_{\hat{P}%
_{j}}(m,\vec{n}),  \label{AI8A}
\end{equation}
the kernel of the star-product being given by Eq. (\ref{eq.26a}). One sees
that, in the probability representation, the process of measurement, both
with von Neumann instruments and with POVM, is described by a map of points
in the simplex.

\section{Time evolution of quantum states and superoperators}

In the standard representation of quantum mechanics, states (state vectors $%
\mid \psi ,t\rangle $ or density operators $\hat{\rho}(t)$) of closed system
evolve according to unitary change 
\begin{equation}
\mid \psi ,t\rangle =\hat{U}(t)\mid \psi ,0\rangle ,  \label{V1}
\end{equation}
or 
\begin{equation}
\hat{\rho}(t)=\hat{U}(t)\hat{\rho}(0)\hat{U}^{\dagger }(t).  \label{V2}
\end{equation}
This evolution is a solution to Schr\"{o}dinger or von Neumann equations 
\begin{equation}
\frac{\partial }{\partial t}\hat{\rho}(t)+i[\hat{H},\rho (t)]=0,  \label{V3}
\end{equation}
$\hat{H}$ being the Hamiltonian of the system.

The evolution can be cast into operator symbol form.

Let $f_{A}(\mathbf{{x})}$ be the symbol of an operator $\hat{A}$. We do not
specify at the moment what kind of symbols are used, considering them as
generic ones with quantizer--dequantizer pair $\hat{D}(\mathbf{{x})}$, $\hat{%
U}(\mathbf{{x})}$. Then, the operator equation (\ref{V3}) for density
operator reads 
\begin{equation}
\frac{\partial }{\partial t}w_{\rho }(\mathbf{x},t)+i\Big(f_{H}(\mathbf{x}%
,t)\star w_{\rho }(\mathbf{x},t)-w_{\rho }(\mathbf{x},t)\star f_{H}(\mathbf{x%
},t)\Big)=0.  \label{V4}
\end{equation}
We denote the symbol of the density operator $\hat{\rho}(t)$ by $w_{\rho }(%
\mathbf{x},t)$. The solution of Eq. (\ref{V4}) has a form corresponding to (%
\ref{V2}) 
\begin{equation}
w_{\rho }(\mathbf{x},t)=f_{U}(\mathbf{x},t)\star w_{\rho }(\mathbf{x}%
,0)\star f_{U^{\dagger }}(\mathbf{x},t).  \label{V5}
\end{equation}
One can rewrite the solution (\ref{V2}) as a superoperator $L$ acting in a
linear space of operators, namely, 
\begin{equation}
\hat{\rho}(t)=L(t)\hat{\rho}(0).  \label{V6}
\end{equation}
In matrix form, Eq. (\ref{V6}) reads 
\begin{equation}
\hat{\rho}(t)_{\alpha \beta }=\sum_{\gamma \delta }L(t)_{\alpha \beta
\,\gamma \delta }\hat{\rho}(0)_{\gamma \delta }.  \label{V7}
\end{equation}
For unitary evolution, the superoperator is expressed in terms of unitary
matrix $U(t)$ as a tensor product 
\begin{equation}
L(t)_{\alpha \beta \,\gamma \delta }=U(t)_{\alpha \beta }\otimes
U^{*}(t)_{\gamma \delta }.  \label{V8}
\end{equation}
One rewrites the solution (\ref{V5}) for the symbol introducing the
propagator 
\begin{equation}
w_{\rho }(\mathbf{x},t)=\int \Pi (\mathbf{x},\mathbf{y},t)w_{\rho }(\mathbf{y%
},0)\,d\mathbf{y}.  \label{V9}
\end{equation}
For the unitary evolution (\ref{V2}), the propagator reads 
\begin{equation}
\Pi (\mathbf{x}_{1},\mathbf{x}_{2},t)=\int k(\mathbf{y}_{1},\mathbf{x}_{2},%
\mathbf{y}_{2},\mathbf{x}_{1})f_{U(t)}(\mathbf{y}_{1})f_{U^{\dagger }(t)}(%
\mathbf{y}_{2})\,d\mathbf{y}_{1}\,d\mathbf{y}_{2},  \label{V10}
\end{equation}
where 
\[
k(\mathbf{y}_{1},\mathbf{x}_{2},\mathbf{y}_{2},\mathbf{x}_{1})=\int K(%
\mathbf{y}_{1},\mathbf{x}_{2},\mathbf{x}_{3})K(\mathbf{x}_{3},\mathbf{y}_{2},%
\mathbf{x}_{1})\,d\mathbf{x}_3. 
\]
The kernels under the integral are given by Eq. (\ref{eq.26}).

In the case of superoperators describing the evolution of an open system 
\cite{Sud61,Kraus} 
\begin{equation}
\hat{\rho}(0)\rightarrow \hat{\rho}(t)=\sum_{s}\hat{V}_{s}(t)\hat{\rho}(0)%
\hat{V}_{s}^{\dagger }(t),\qquad \sum_{s}\hat{V}_{s}^{\dagger }(t)\hat{V}%
_{s}(t)=1,  \label{Z1}
\end{equation}
the propagator reads 
\begin{equation}
\Pi (\mathbf{x},\mathbf{y},t)=\int \sum_{s}f_{V(t)}^{(s)}(\mathbf{y}%
_{1})f_{V^{\dagger }(t)}^{(s)}(\mathbf{y}_{2})k(\mathbf{y}_{1},\mathbf{y},%
\mathbf{y}_{2},\mathbf{x})\,d\mathbf{y}_{1}\,d\mathbf{y}_{2}.  \label{Z2}
\end{equation}
The propagator corresponds to a superoperator, which in matrix form reads 
\begin{equation}
\Big(L(t)\Big)_{\alpha \beta \,\gamma \delta }=\sum_{s}(V_{s})_{\alpha \beta
}\otimes (V_{s}^{*})_{\gamma \delta }.  \label{Z3}
\end{equation}
For the case of continuous variables and symplectic tomograms, Eq. (\ref{V4}%
) takes the form of a deformed Boltzman equation for the probability
distribution.

For unitary spin tomograms, one has 
\begin{equation}
w_{\rho }(m,u,t)=w_{\rho }\left( m,U^{\dagger }(t)u,0\right) ,  \label{Z4}
\end{equation}
the unitary evolution matrix being determined by an Hamiltonian matrix 
\begin{equation}
U(t)=e^{-itH}.  \label{Z5}
\end{equation}
This means that the unitary spin tomogram, a function on the unitary group,
evolves according to the regular representation of the unitary group. This
means that the partial differential equation for the infinitesimal action is
the standard equation for matrix elements of the regular representation,
that is, 
\[
i\frac{\partial }{\partial t}w(m,u,t)=\sum_{ik}\Big(H_{ik}\hat{L}_{ik}(u)%
\Big)
w(m,u,t),
\]
$H_{ik}$ being the Hamiltonian hermitian matrix and $\hat{L}_{ik}(u)$ the
infinitesimal hermitian first-order differential operators of the left
regular representation of the unitary group in the chosen group
parametrization.

\section{Examples of quantum channels}

In this section, we consider the unitary spin representation of some typical
quantum channels.

\subsection{Depolarizing channel}

Consider bit flip, phase flip and both with equal probability 
\[
\begin{array}{l}
\mid \psi \rangle \rightarrow \sigma _{1}\mid \psi \rangle =\left( 
\begin{array}{ll}
0 & 1 \\ 
1 & 0
\end{array}
\right) \mid \psi \rangle, \\ 
\mid \psi \rangle \rightarrow \sigma _{3}\mid \psi \rangle =\left( 
\begin{array}{ll}
1 & 0 \\ 
0 & -1
\end{array}
\right) \mid \psi \rangle , \\ 
\mid \psi \rangle \rightarrow \sigma _{2}\mid \psi \rangle =\left( 
\begin{array}{ll}
0 & -i \\ 
i & 0
\end{array}
\right) \mid \psi \rangle.
\end{array}
\]
The Kraus representation is 
\[
\rho _{0}\rightarrow \rho =\left( 1-p\right) \rho +\frac{p}{3}\left( \sigma
_{1}\rho \sigma _{1}+\sigma _{2}\rho \sigma _{2}+\sigma _{3}\rho \sigma
_{3}\right). 
\]
For the unitary spin symbol representation, choosing the axis one has 
\[
\rho _{0}=\frac{1}{2}(1+\sigma _{3}) 
\]
and with an arbitrary unitary group element 
\[
u=\cos \frac{\theta }{2}-i\sigma\cdot n\sin \frac{\theta }{2} 
\]
one obtains 
\[
\begin{array}{lll}
w\left( +,u\right) & = & \frac{1}{2}\left\{ 1+\left( 1-\frac{4}{3}p\right)
\left( \cos ^{2}\frac{\theta }{2}+\left( 2n_{3}^{2}-1\right) \sin ^{2}\frac{%
\theta }{2}\right) \right\}, \\ 
w\left( -,u\right) & = & \frac{1}{2}\left\{ 1-\left( 1-\frac{4}{3}p\right)
\left( \cos ^{2}\frac{\theta }{2}+\left( 2n_{3}^{2}-1\right) \sin ^{2}\frac{%
\theta }{2}\right) \right\}.
\end{array}
\]
When $p\rightarrow 1$ the image in the simplex contracts to a segment
between $\left( \frac{2}{3},\frac{1}{3}\right) $ and $\left( \frac{1}{3},%
\frac{2}{3}\right) $.

\subsection{Phase-damping channel}

\[
\begin{array}{l}
\mid 0\rangle \mid 0\rangle _{E}\rightarrow \sqrt{1-p}\mid 0\rangle \mid
0\rangle _{E}+\sqrt{p}\mid 0\rangle \mid 1\rangle _{E}, \\ 
\mid 1\rangle \mid 0\rangle _{E}\rightarrow \sqrt{1-p}\mid 1\rangle \mid
0\rangle _{E}+\sqrt{p}\mid 1\rangle \mid 2\rangle _{E}.
\end{array}
\]
The Kraus representation is 
\[
\rho _{0}\rightarrow \rho =\sum_{\mu }K_{\mu }\rho _{0}K_{\mu }^{\dagger }, 
\]
with 
\begin{eqnarray*}
K_{0} &=&\sqrt{1-p}\left( 
\begin{array}{ll}
1 & 0 \\ 
0 & 0
\end{array}
\right), \qquad K_{1}=\sqrt{p}\left( 
\begin{array}{ll}
1 & 0 \\ 
0 & 0
\end{array}
\right), \\
\qquad K_{2} &=&\sqrt{p}\left( 
\begin{array}{ll}
0 & 0 \\ 
0 & 1
\end{array}
\right).
\end{eqnarray*}

For the unitary spin symbol representation, consider the example 
\[
\mid \psi \rangle =\frac{1}{\sqrt{2}}\left( \mid 0\rangle +\mid 1\rangle
\right), 
\]
\[
\rho _{0}=\left( 
\begin{array}{ll}
\frac{1}{2} & \frac{1}{2} \\ 
\frac{1}{2} & \frac{1}{2}
\end{array}
\right). 
\]
Then 
\[
\begin{array}{lll}
w\left( +,u\right) & = & \frac{1}{2}\left\{ 1+2\left( 1-p\right) \sin \frac{%
\theta }{2}\left( n_{2}\cos \frac{\theta }{2}+n_{1}n_{3}\sin \frac{\theta }{2%
}\right) \right\}, \\ 
w\left( -,u\right) & = & \frac{1}{2}\left\{ 1-2\left( 1-p\right) \sin \frac{%
\theta }{2}\left( n_{2}\cos \frac{\theta }{2}+n_{1}n_{3}\sin \frac{\theta }{2%
}\right) \right\}.
\end{array}
\]
and when $p\rightarrow 1$ the image in the simplex contracts to a point.

\subsection{Amplitude damping channel}

\[
\begin{array}{l}
\left| 0\right\rangle \left| 0\right\rangle _{E}\rightarrow \left|
0\right\rangle \left| 0\right\rangle _{E}, \\ 
\left| 1\right\rangle \left| 0\right\rangle _{E}\rightarrow \sqrt{1-p}\left|
1\right\rangle \left| 0\right\rangle _{E}+\sqrt{p}\left| 0\right\rangle
\left| 1\right\rangle _{E}.
\end{array}
\]
The Kraus representation is 
\[
\rho _{0}\rightarrow \rho =\sum_{\mu }K_{\mu }\rho _{0}K_{\mu }^{\dagger }, 
\]
with 
\[
K_{0}=\left( 
\begin{array}{ll}
1 & 0 \\ 
0 & \sqrt{1-p}
\end{array}
\right), \qquad K_{1}=\left( 
\begin{array}{ll}
0 & \sqrt{p} \\ 
0 & 0
\end{array}
\right). 
\]

For the unitary spin representation, consider an excited initial state 
\[
\mid \psi \rangle =\mid 1\rangle , 
\]
\[
\rho _{0}=\left( 
\begin{array}{ll}
0 & 0 \\ 
0 & 1
\end{array}
\right) . 
\]
Then 
\[
\begin{array}{lll}
w\left( +,u\right) & = & p\cos ^{2}\frac{\theta }{2}+\left(
pn_{3}^{2}+\left( 1-p\right) \left( 1-n_{3}^{2}\right) \right) \sin ^{2}%
\frac{\theta }{2}\,, \\ 
w\left( -,u\right) & = & \left( 1-p\right) \cos ^{2}\frac{\theta }{2}+\left(
\left( 1-p\right) n_{3}^{2}+p\left( 1-n_{3}^{2}\right) \right) \sin ^{2}%
\frac{\theta }{2}\,.
\end{array}
\]
When $p$ varies from $0$ to $1$, the image in the simplex first contracts to
a point $\left( p=\frac{1}{2}\right) $ and then expands again to the whole
simplex when $p\rightarrow 1$.

The operator symbols being functions on the rotation or unitary groups are
highly redudant descriptions of qudit states. As expected from the number of
independent parameters in the density matrix, also here $\left(
d^{2}-1\right) $ numbers are enough to characterize a $d-$dimensional qudit.
This is easy to check. Consider the operator symbol (\ref{A86}) for an
arbitrary $d-$dimensional density matrix $\rho $. A general $\rho $ may be
diagonalized by of $d\left( d-1\right) $ independent unitary transformations
and this, together with the $\left( d-1\right) $ independent diagonal
elements, gives the desired result.

Alternatively we may consider $\left( d+1\right) $ independent elements of
the unitary group and compute the associated operator symbols. Then, the
qudit state would be described by their diagonal elements. Therefore, a
discrete quantum state (qudit) is coded by $\left( d+1\right) $ probability
distributions.

For each $u$ in the group, the elements in the operator symbol $w(\left\{
m_{i}\right\} ,u)$ are the probabilities to obtain the values $\left\{
m_{i}\right\} $ in a measurement of the quantum state $\rho $ by an
apparatus oriented along $u$. Therefore the problem of reconstructing the
state $\rho $ from the set of $\left( d+1\right) \left( d-1\right) $
operator symbol elements is identical to the reconstruction of the density
matrix of a spin through Stern-Gerlach experiments, already discussed in the
literature\cite{Newton} \cite{Weigert1} \cite{Weigert2}.

\section{Entropies}

\subsection{Operator symbol entropies}

The tomographic operator symbols satisfy 
\[
\sum_{\left\{ m_{i}\right\} }w(\left\{ m_{i}\right\} ,u)=1, 
\]
therefore, they are probability distributions $\forall u$.

One defines the \textit{operator symbol entropy} by 
\[
H_{u}=-\sum_{\left\{ m_{i}\right\} }w(\left\{ m_{i}\right\} ,u)\ln w(\left\{
m_{i}\right\} ,u) 
\]
and the \textit{operator symbol R\'{e}nyi entropies} by 
\[
R_{u}=\frac{1}{1-q}\ln \left( \sum_{\left\{ m_{i}\right\} }w(\left\{
m_{i}\right\} ,u)^{q}\right). 
\]

Likewise, we may define the \textit{operator symbol relative q-entropy} by 
\begin{equation}
H_{q}(w_{1}(u)|w_{2}(u))=-\sum_{\left\{ m_{i}\right\} }w_{1}(\left\{
m_{i}\right\} ,u)\ln _{q}\frac{w_{2}(\left\{ m_{i}\right\} ,u)}{%
w_{1}(\left\{ m_{i}\right\} ,u)},  \label{ZZZ3}
\end{equation}
with 
\begin{equation}
\ln _{q}x=\frac{x^{1-q}-1}{1-q}\,,\quad x>0,\quad q>0,\quad \ln
_{q\rightarrow 1}x=\ln x.  \label{ZZZ2}
\end{equation}
Because the operator symbols $w\left( \left\{ m_{i}\right\} ,u\right) $ are
probability distributions, they inherit all the known properties of
nonnegativity, additivity, joint convexity, etc. of classical information
theory.

The relation of the operator symbol entropies to the von Neumann and the
quantum R\'{e}nyi entropies is given by the following

\begin{teo}
The von Neumann $S$ and the quantum R\'{e}nyi $S_{q}$ entropies are the
minimum on the unitary group of $H_{u}$ and $R_{u}$.
\end{teo}

\textit{Proof} : From 
\begin{equation}
w(\left\{ m_{i}\right\} ,u)=\langle \left\{ m_{i}\right\} \mid u^{\dagger
}\rho u\mid \left\{ m_{i}\right\} \rangle  \label{tt1}
\end{equation}
there is a $u^{*}$ such that $u^{*\dagger }\rho u^{*}$ is a diagonal matrix $%
\left\{ \lambda _{1},\lambda _{2},\ldots\lambda _{n}\right\} $. Then 
\[
H_{u^{*}}=-\mbox{Tr}\,\rho \ln \rho =S=\mbox{von\,\,Neumann\,\,
entropy} 
\]
For any other $u$, the diagonal elements in (\ref{tt1}) are convex linear
combinations of $\left\{ \lambda _{1},\lambda _{2},...\lambda _{n}\right\} $%
. By convexity of $w(\left\{ m_{i}\right\} ,u)\ln w(\left\{ m_{i}\right\}
,u) $ the result follows for the von Neumann entropy.

The operator symbol R\'{e}nyi entropy is not a sum of concave functions.
However, the following function is,
\[
T_{u}=-\sum_{\left\{ m_{i}\right\} }w(\left\{ m_{i}\right\} ,u)^{q}\ln
_{q}w(\left\{ m_{i}\right\} ,u)
\]
called the Tsallis entropy\cite{tsallis} and related to the R\'{e}nyi\cite
{renyi} entropy by 
\begin{equation}
R_{u}=\frac{1}{1-q}\ln \Big( 1+\left( 1-q\right) T_{u}\Big).  \label{tt2}
\end{equation}

The minimum result now applies to $T_{u}$ by concavity and then one checks
from (\ref{tt2}) that it also holds for $R_{u}$. Therefore $%
R_{u^{*}}=R_{u}\min $ coincides with the quantum R\'{e}nyi entropy 
\[
S_{q}=\frac{1}{1-q}\ln (\mbox{Tr}\,\rho ^{q}).
\]
{\raggedright $\Box $}

The entropy $H_{u}$ varies from the minimum, which is von Neumann entropy,
to a maximum for the most random distribution. For each given state $\rho $,
one can also define the \textit{integral entropies}
\[
H_{\rho }=\int H_{u}\,du,\qquad R_{\rho }=\int R(u)\,du,\qquad H_{q}(\rho
_{1},\rho _{2})=\int H_{q}(w_{1}(u)|w_{2}(u))\,du,
\]
where $du$ is the invariant Haar measure on unitary group.

In some cases, the properties of the von Neumann entropy may be derived as
simple consequences of the classical-like properties of the operator symbol
entropies. For example:

\textbf{Subadditivity:} $S_{12}\leq S_{1}+S_{2}$

Consider a two-partite system with density matrix $\rho _{12}$%
\[
w(m_{1},m_{2},u)=\langle m_{1}m_{2}\mid u^{\dagger }\rho _{12}u\mid
m_{1}m_{2}\rangle , 
\]
$u$ being a $(2j_{1}+1)(2j_{2}+1)\times (2j_{1}+1)(2j_{2}+1)$ unitary matrix 
\[
H_{u}\left( 12\right) =-\sum_{m_{1}m_{2}}w(m_{1},m_{2},u)\ln
w(m_{1},m_{2},u) 
\]

From the reduced symbols and density matrices 
\[
w(m_{1},u)=\sum_{m_{2}}w(m_{1},m_{2},u), 
\]
\[
(\rho _{1})_{m_{1}m_{1}^{\prime }}=\sum_{m_{2}}(\rho
_{12})_{m_{1}m_{2}m_{1}^{\prime }m_{2}}, 
\]
one writes the reduced symbol entropies 
\[
H_{u}(1)=-\sum_{m_{1}}w(m_{1},u)\ln \,w(m_{1},u), 
\]
\[
H_{u}(2)=-\sum_{m_{2}}w(m_{2},u)\ln \,w(m_{2},u). 
\]
For each fixed $u$, the tomographic symbols are ordinary probability
distributions. Therefore, by the subadditivity of classical entropy, 
\[
H_{u}(12)\leq H_{u}(1)+H_{u}(2). 
\]
In particular, this is true for the group element in $u^{*}\in
U_{j_{1}}\otimes U_{j2}$ that diagonalizes the reduced density matrices $%
\rho _{1}$ and $\rho _{2}$. Therefore, 
\[
H_{u^{*}}(12)\leq S_{1}+S_{2}. 
\]
But, by the minimum property, 
\[
S_{12}\leq S_{1}+S_{2}. 
\]

Thus subadditivity for the von Neumann entropy is a consequence of
subadditivity for the operator symbol entropies. 

The situation concerning strong subadditivity is different. Strong
subadditivity also holds, of course, for the operator symbol entropies for
any $u$%
\begin{equation}
H_{u}(123)+H_{u}(2)\leq H_{u}(12)+H_{u}(23),  \label{tt3}
\end{equation}
but the corresponding relation for the von Neumann entropy is not a direct
consequence of (\ref{tt3}). The strong subadditivity\cite{Ruskai} for the
von Neumann entropy 
\begin{equation}
S\left( 123\right) +S\left( 2\right) \leq S\left( 12\right) +S\left(
23\right)   \label{tt4}
\end{equation}
expressed in operator symbol entropies would be 
\[
H_{u_{1}^{*}}(123)+H_{u_{2}^{*}}(2)\leq H_{u_{3}^{*}}(12)+H_{u_{4}^{*}}(23)
\]
where $u_{1}^{*}\in U\left( 123\right) ,u_{2}^{*}\in 1\otimes U(2)\otimes
1,u_{3}^{*}\in U\left( 12\right) \otimes 1,u_{4}^{*}\in 1\otimes U(23)$ are
the different group elements that diagonalize the respective subspaces.
Therefore strong subadditivity for the von Neumann entropy (\ref{tt4}) and
strong subadditivity for the operator symbol entropies (\ref{tt3}) are
independent properties. On the other hand, because of the invertible
relation (\ref{equation29}) between the operator symbols and the density
matrix, Eq.(\ref{tt3}) contains in fact a family of new inequalities for
functionals of the density matrix.

\section{Conclusions}

To conclude we summarize the main results of this work :

(i) A unified formulation for an operator symbol formulation of standard
quantum theory.

(ii) A (spin) operator symbol framework to deal with quantum information
problems.

(iii) Evolution equations for qudit operator symbols are written in the form
of first-order partial differential equations with generators describing the
left regular representation of the unitary group.

(iv) Measurements are discussed in the operator-symbol representation of
qudits.

(v) A geometric interpretation of (spin) operator symbols of qudit states as
maps of the unitary group to the simplex.

(vi) In view of the probability nature of the operator symbols, the
corresponding entropies inherit the properties of classical information
theory. Some of the properties of the von Neumann entropy and quantum R\'{e}%
nyi entropy are direct consequences of these properties. On the other hand
the properties of the operator symbol entropies also imply new relations for
functionals of the density matrix.

\end{document}